\documentclass[referee]{raa}
\usepackage{graphicx,times}
\usepackage{natbib}
\usepackage{amssymb,amsmath}
\bibpunct{(}{)}{;}{a}{}{,}

\usepackage[a4paper=true,dvipdfm=true,pagebackref=true]{hyperref}
\hypersetup{pdftitle = The title of my PDF, pdfauthor = My name, pdfsubject= The subject, pdfkeywords = keyword1 keyword2 keyword3}
\hypersetup{colorlinks = true, linkcolor = green, anchorcolor = red, citecolor = blue, filecolor = red, pagecolor = red, urlcolor = red}

\begin{document}

   \title{Ultra-Low Noise L-Band Cryogenic Astronomical Receiver for FAST Telescope
$^*$
\footnotetext{\small $*$ Supported by National Key R\&D Program of China and National Natural Science Foundation of China.}
}

 \volnopage{ {\bf 20XX} Vol.\ {\bf X} No. {\bf XX}, 000--000}
   \setcounter{page}{1}

   \author{Hong-Fei Liu\inst{1,2}, Chuan He\inst{3}, Jin Wang\inst{4}, Peng Jiang\inst{1,2},  Sheng-Wang Wang\inst{3}, Yang Wu\inst{4}, Hang Zhang\inst{1,2}, Jin-You Song\inst{1,2}, Xiang-Wei Shi\inst{1,2} and Ming-Lei Guo\inst{1,2}
   }

   \institute{ National Astronomical Observatories, Chinese Academy of Sciences (CAS), Beijing 100012,
China; {\it lhf@bao.ac.cn}\\
        \and
             CAS Key Laboratory of FAST, National Astronomical Observatories, Chinese Academy of Sciences,
             Beijing 100012, China\\
	\and
The 16th Research Institute of China Electronic Technology Group Corporation, Hefei 230043, China\\
\and
The 54th Research Institute of CETC,
 Shijiazhuang  050081, China\\
\vs \no
   {\small Received 2021 March 12; accepted 2021 March 12}
}

\abstract{This paper presents an ultra-low noise L-band radio astronomical cryogenic receiver for FAST telescope. The development of key low noise microwave parts of Coupling-LNA and conical quad-ridge OMT and reasonable system integration achieve outstanding performance of receiver. It covers the frequency range of 1.2 GHz to 1.8 GHz. Novel cryogenic Coupling-LNAs with low noise, large return loss, high dynamic range and the function of coupling calibration signals are developed for the proposed receiver. Amplification and coupling function circuits are integrated as a single Coupling-LNA with full noise temperature of 4 K at the physical temperature of 15 K. And its return loss is more than 18 dB, and output 1 dB compression power is +5 dBm. A cryogenic dewar is fabricated to provide 55 K and 15 K cryogenic environment for OMT and Coupling-LNAs, respectively. The receiver's system noise temperature is below 9 K referred to feed aperture plane. Benefiting from optimal design and precise mechanical treatment, good scattering performance of OMT and equalized radiation patterns of horn are achieved with an antenna efficiency above 75\%.
}

   \authorrunning{H.-F. Liu et al. }            
   \titlerunning{Low noise L band cryogenic receiver for FAST}  
   \maketitle

%
\section{Introduction}           
\label{sect:intro}

Five-hundred-meter Aperture Spherical radio Telescope (FAST) is the largest single aperture radio telescope on earth. The primary reflector of FAST is a spherical crown with 500 m aperture. When FAST is set to point at a certain astronomical source, the corresponding reflecting units will be pulled or pushed to form a 300 m aperture paraboloid with specific orientation \citep{Nan2006}. Such a huge receiving area (A) requires the receiver noise (T) to be as low as possible, so as to achieve ultra-high detecting sensitivity of telescope in view of basic sensitivity expression of A/T. Seven receivers covering the band from 0.07 GHz to 3 GHz were built for FAST telescope. Wherein, L-band is the most important frequency band for FAST in view of relatively low sky noise and relatively high antenna efficiency. There are two sets of L-band receivers for FAST. Wherein, 19-beam receiver has been running on telescope since 2018 \citep{Jiang2020}, which covers 1.05 GHz to 1.45 GHz. There is no doubt that the 19-beam receiver has higher sky-surveying efficiency in view of the beam numbers 19 times more than common single pixel receivers. While, limited by the size of receiver cryostat, it is hard to construct the horns and orthogonal mode transducers (OMT) operating in wider frequency band for 19-beam receiver. The other one is the proposed L-band cryogenic receiver in this paper, which has one beam, however, covers a wider frequency band of 1.2 GHz to 1.8 GHz.

Based on the physical optics features of FAST telescope \citep{Nan2006}, optimal design and high precision machining for horn and OMT of the proposed receiver are made to ensure good equalization of radiation pattern and excellent scattering performance across 600 MHz passband.

A Gifford-McMahon (GM) Helium cooling system, including vacuum dewar, Helium cold head and compressor, is developed to cool receiver frontend low noise microwave components and resultantly reduce receiver overall noise. Wherein, the first stage of cold head is used to cool OMT to an average physical temperature of 55 K (35 K at bottom and 75 K at aperture). In most applications, directional couplers and low noise amplifiers (LNA) as two independent components are individually cooled to about 15 K by the second stage of cold head. Directional couplers for injecting calibration signals into receiver signal channels are cascaded in front of LNAs. Low Noise Factory, Inc. (LNF) and Cosmic Microwave Technology, Inc. (CMT) offer the best performing commercial cryogenic LNAs, wherein, LNF claims state-of-the-art LNA of Indium Phosphide (InP) based LNC0.6-2A \cite{LNC0.6-2A}
with 1 K noise at a physical temperature of 4 K in L band. While, its input return loss in 1.72 GHz to 1.8 GHz is less than 10 dB, which may cause some ripples of passband response for the proposed receiver. In addition, 1 K noise is achieved at the physical temperature of 4 K, which will increase a little in this receiver with the cryogenic temperature of 15 K. The trade-off design between noise and power matching is the primary topic in LNA development to ensure low noise and low reflection, and a usable cryogenic LNA applied in astronomical receiver should have an input return loss of more than 10 dB at least. CMT offers cryogenic LNA of CITLF3 with Silicon-Germanium (SiGe) based Heterojunction Bipolar Transistor (HBT) transistors with input return loss of 10 dB - 12 dB, while its noise temperature increases to 3 K - 4.2 K at the ambient temperature of 12.5 K from 1.2 GHz to 1.8 GHz. A typical commercial cryogenic-usable directional coupler, such as C6600 from Werlatone \cite{C6600} has a noise temperature of about 1 K at the physical temperature of 15 K. This means that the cascaded noise of coupler and CITLF3 are about 4 K - 5.2 K. The dynamic ranges of LNC0.6-2A and CITLF3 are -10 dBm and -5 dBm referred to output, respectively, both of which are quite low. For the application of this proposed receiver, a kind of L-band cryogenic Coupling-LNA with low noise, large return loss and high dynamic range has been developed. Novel design in integrated cryogenic LNA and coupler for reducing full noise of component is implemented. That is, the coupling and amplification function circuits are designed as a single component of Coupling-LNA with one pair of input/output passive circuits and connectors. Which reduces passive elements of both components and resultant the overall noise of Coupling-LNA. While, the trade-off design in noise vs. power matching, high gain, stabilization and high dynamic range for this cryogenic Coupling-LNA becomes more complex. At the physical temperature of 15 K, the noise temperature of the fabricated Coupling-LNA is from 3.7 K to 4.4 K, and input return loss is greater than 18 dB. Meanwhile, its 1 dB compression power referred to output is +5 dBm.

Hermetic properties of large rectangle dewar is another key in the fabrication of cryogenic receiver. In order to achieve good scattering performance in L band, a long OMT (728.3 mm in length) is adopted, which requires a bigger size of cryogenic dewar. Moreover, most traditional cryogenic receivers have cylindrical dewar structure to achieve good hermeticity. While, cylindrical structure makes it difficult to open the cavity in receiver cabin of telescope when fixing or replacing the components inside cavity. For the proposed receiver, a rectangular cryogenic dewar is built, and its front panel can be removed easily when maintenance is required.  In view of above characteristics of dewar, reasonable hermeticity design and precision thermal analysis are implemented. The vacuum degree of $10^{-5}$ Pa, the second stage of 15 K and the first stage of 55 K cold head temperatures are achieved inside the fabricated dewar.

\begin{table}
\caption{Comparison of main performances of similar receivers worldwide}
\centering
\begin{tabular}{|c|c|c|c|}
\hline
No.	& Telescopes with similar receivers & Operating Frequencies & Noise Temperatures\\\hline
1 & Sheshan \cite{Woestenburg} & 1.6 GHz - 1.74 GHz & 15 K\\\hline
2 & SRT \cite{Navarrini} & 1.3 GHz - 1.8 GHz & 10 K - 13 K\\\hline
3 & Effelsberg \citep{Hachenberg1973} & 1.2 GHz - 1.8 GHz & 10 K\\\hline
4 & GBT \citep{Bolli2019} & 1 GHz - 2 GHz & 8 K - 10 K\\\hline
5 & This paper & 1.2 GHz - 1.8 GHz & less than 9 K\\\hline

\end{tabular}
\label{T1}
\end{table}

Above key points ensure the noise temperature of receiver system are below 9 K referred to feed aperture plane, and it is an ultra-low noise L-band receiver. Compared with the other high-performance astronomical receivers in similar band worldwide (shown in Table \ref{T1}), the highlights of the proposed receiver are the development of key low noise microwave components: Coupling-LNA and OMT. The Coupling-LNA has the characteristics of low noise, large return loss, high dynamic range and the function of coupling calibration signals. Therefore, it has good application value. Conical quad-ridge OMT eliminates the trapped mode resonances of traditional OMT with square waveguide, and resultantly ensures good radiation performance.

\section{DESIGN AND FABRICATION OF RECEIVER}
\label{sect:Obs}

\begin{figure}
   \centering
   \renewcommand\thefigure{1}
   \includegraphics[width=14.0cm, angle=0]{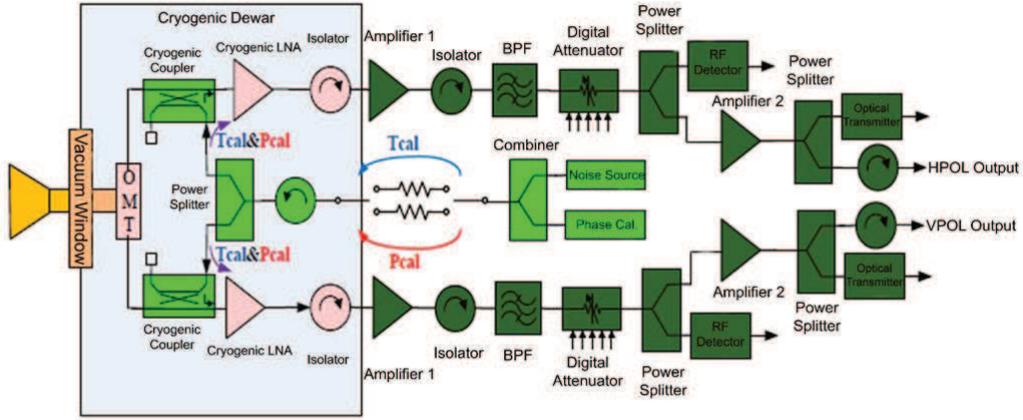}
   \caption{Block diagram of FAST L-band cryogenic receiver. Corrugated horn, OMT, cryogenic dewar, cryogenic microwave components, warm microwave and calibrating signals injection units are shown. There are two signal channels corresponding to H and V polarization of receiver's horn.}
   \label{Fig1}
   \end{figure}

\begin{figure}
   \centering
   \renewcommand\thefigure{2}
   \includegraphics[width=12.0cm, angle=0]{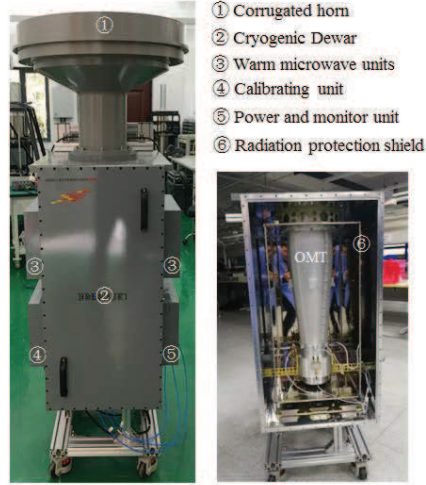}
   \caption{ Photographs of the assembled receiver (left) and receiver dewar with front panel removed (right).  }
   \label{Fig2}
   \end{figure}

The block diagram and physical image of the proposed receiver are shown in Fig.\ref{Fig1} and Fig.\ref{Fig2}, respectively. The receiver consists of feed network, cryogenic cooling unit, warm microwave units, calibrating signals injection unit and system power supply and monitoring unit. Based on the noise theory for a cascaded system described in Equation (\ref{eq:1}), the noise contribution of microwave parts in front of first stage LNA in signal chain must be as low as possible to achieve overall low noise performance.
\begin{equation}
  T_{e} = T_{e1}+\frac{T_{e2}}{G_{1}}+\frac{T_{e3}}{G_{1}G_{2}}+...
\label{eq:1}
\end{equation}

Corresponding to receiver system, $T_e$ is the equivalent noise temperature of receiver system; $T_{e1}$ is mainly contributed by horn, vacuum window, OMT, cryogenic coaxial cables, cryogenic directional coupler and cryogenic LNA; $T_{e2}$ and $T_{e3}$ are noise temperature of backstage components of cryogenic LNA. And $G_1$ and $G_2$ are power gain of cryogenic LNA and RF amplifier.

System design analysis on ambient temperature (${\rm T_{amb}}$), gain in logarithm (${\rm G_{dB}}$), gain in linearity (${\rm G_{Lin}}$), noise temperature (${\rm T_e}$), noise contribution to system (${\rm \bigtriangleup T}$) and output 1 dB compression point (${\rm P_{out-1\ dB}}$) of each microwave parts in receiver chain is shown in Table \ref{T2}. Wherein, the calculated system specification are as follows: gain is 65.15 dB; system output 1 dB compression point is 10 dBm; system noise is 8.97 K referred to feed aperture plane (7.59 K referred to vacuum window).

\begin{table}
\caption{Main specifications breakdown of receiver microwave chain}
\centering
\begin{tabular}{|c|c|c|c|c|c|c|}
\hline
Microwave parts	   & ${\rm T_{amb}}$ & ${\rm G_{dB}}$   & ${\rm G_{Lin}}$  & ${\rm T_e}$   & ${\rm \bigtriangleup T}$   & ${\rm P_{out-1\ dB}}$                          \\\hline
            	   & K    & dB    &       & K    & K    &  dBm        \\\hline
Feed	           & 300  & -0.02 & 0.995 & 1.38 & 1.38 &             \\\hline
Vacuum window	   &300   &-0.01  &0.998  &0.69  &0.69  &	          \\\hline
OMT	               &55	  &-0.12  &0.973  &1.54  &1.55  &	          \\\hline
Cryogenic cable	   &15	  &-0.1   &0.977  &0.35  &0.36  &	          \\\hline
Noise source	   &	  &	      &       &0.45  &0.48  &	          \\\hline
Coupling-LNA	   &15	  &30	  &1000   &4	 &4.24  &5	          \\\hline
Insulation cable   &150	  &-1	  &0.794  &38.84 &0.04  &              \\\hline
Hermetic connector &300	  &-0.5	  &0.891  &36.61 &0.05  &              \\\hline
Amplifier1	       &300	  &30	  &1000	  &120	 &0.18	&5             \\\hline
Isolator	       &300	  &-0.6	  &0.871  &44.45 &0.00  &	           \\\hline
Passband filter	   &300	  &-1	  &0.794  &77.68 &0.00  &	            \\\hline
Digital attenuator &300	  &-1.5	  &0.708  &123.76&0.00  &	            \\\hline
Power splitter	   &300	  &-4	  &0.398  &453.57&0.00	&             \\\hline
Amplifier2	       &300	  &20	  &100	  &650	 &0.00	&15           \\\hline
Power splitter	   &300	  &-4	  &0.398  &453.57&0.00  &	           \\\hline
RFoF	           &300	  &0	  &1	  &91416 &0.01	&10           \\\hline
Warm cables   	   &300	  &-2	  &0.631  &175.47&0.00  &	          \\\hline
System specs	   &      &65.15  &		  &      &8.97	&10           \\\hline

\end{tabular}
\label{T2}
\end{table}

\subsection{Feed network}
\label{sect:Obs}

\begin{figure}
   \centering
   \renewcommand\thefigure{3}
   \includegraphics[width=10.0cm, angle=0]{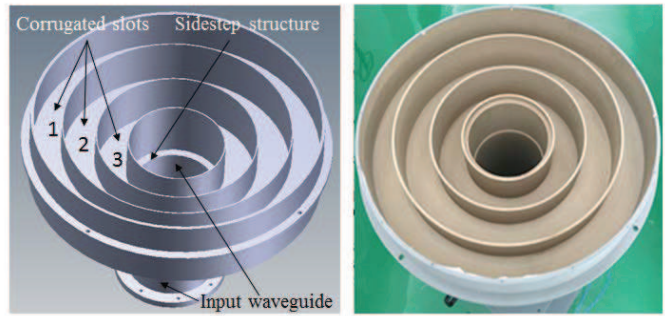}
   \caption{ Simulation model (left) and photograph of corrugated horn (right) of L-band cryogenic receiver. Aluminum with electric conductive oxidation layer on inner walls are employed. The outmost aperture of horn's flare and height of the horn are 678.5 mm and 448.4 mm, respectively. The depth of corrugated slot 1, 2 and 3 are 65 mm, 76 mm and 92 mm, respectively. The length and internal diameter of input waveguide of the horn are 251 mm and 189 mm, respectively.  }
   \label{Fig3}
   \end{figure}

The feed network consists of a horn in atmosphere (about 300 K) and an OMT in vacuum that are electrically interconnected via microwave window. To match main focus reflector of FAST with 0.4621 in the ratio of focal length to aperture, a corrugated horn with $112^{\circ}$ flare angle is designed and fabricated. The optimal simulation model and physical image of the horn are shown in Fig.\ref{Fig3}. The corrugated slots are used to eliminate the E-plane edge currents and resultantly edge diffraction of electromagnetic wave. The surface reactance in the rim of the horn depending on wavelength $\lambda$ and slot depth $d$ can be expressed as:
\begin{equation}
  X \approx 377tan\lbrack\frac{2 \pi d}{\lambda}\rbrack
\label{eq:2}
\end{equation}

If the depth $d$ is between $\lambda/4$ and $\lambda/2$ , the surface reactance will be capacitive and mismatching with free space, and edge radiation will resultantly be repressed. Then, good equalization between E-plane and H-plane radiation patterns will be achieved, and multiple slots with different depths can help over a wide band. The specific depths of slots as well as the other dimensions of horn are figured out by optimal simulation aiming to as high as possible aperture efficiency with a 3D electromagnetic simulator \cite{CST}. The final depth of slots (see Fig.\ref{Fig3}) corresponds to $\lambda_c/3$ , $\lambda_c/2.6$  and $\lambda_c/2.2$, respectively. $\lambda_c$  is the central wavelength of 1.5 GHz. Furthermore, the sidestep structure from input waveguide to corrugated slots is employed to improve matching between circular waveguide and corrugated waveguide.

\begin{figure}
 \centering
 \renewcommand\thefigure{4-1}
   \includegraphics[width=12.0cm, angle=0]{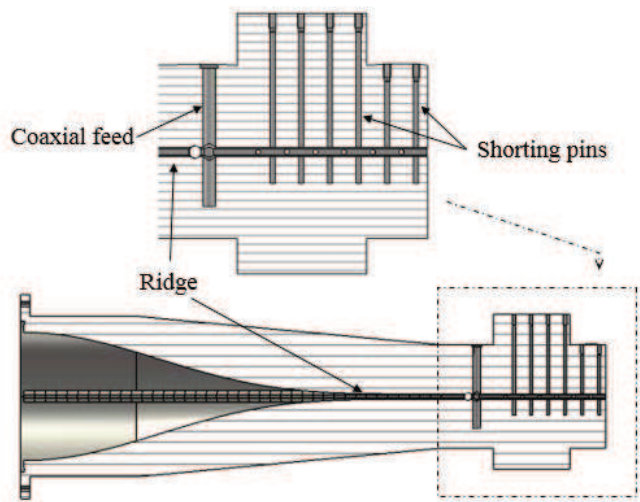}
   \caption{Cross-section of conical quad-ridge OMT of FAST L-band cryogenic receiver. Coaxial feed, shorting pins and ridge can be seen.}
   \label{Fig4-1}
   \end{figure}

\begin{figure}
   \centering
   \renewcommand\thefigure{4-2}
   \includegraphics[width=12.0cm, angle=0]{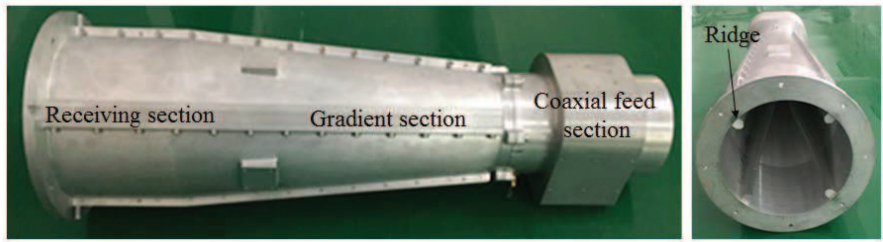}
   \caption{ Photographs of quad-ridge OMT of FAST L-band cryogenic receiver. The length and internal receiving aperture of OMT are 728.3 mm and 189 mm, respectively.  }
   \label{Fig4-2}
   \end{figure}

The waveguide-type horn demonstrates very small dissipative RF losses, so the contribution on full receiver noise temperature is low. Moreover, the thermal load of large horn makes it difficult to be placed inside cryogenic dewar, so the proposed horn works in atmosphere. Following with horn, a quad-ridge OMT is fabricated to transmit and transduce signals from horn to two orthogonal output signals. Traditional quad-ridge OMT \citep{Coutts2011} adopts square waveguide with four ridges in gradient section and requires a square to circular waveguide in receiving section, which will inevitably spark high-order modes and resultantly the trapped mode resonances \citep{Morgan2013}. This paper employed circular waveguide with quad-ridge in gradient section of OMT to eliminate the trapped mode resonances. In addition, the OMT in this paper employs six cylindroid shorting pins to depress reverse radiation of fundamental mode. Compared with traditional rectangular shorting pieces, it helps to improve consistency of return loss between two coaxial feed ports (shown in Fig.\ref{Fig4-1}). To achieve good scattering performance, the length of OMT is optimized to be 728.3 mm. The smooth internal surface of OMT is made by mechanical polishing to achieve low insertion loss. The OMT is installed inside a cryogenic dewar and is cooled to an average physical temperature of 55 K (35 K at bottom and 75 K at aperture) by the first stage of cold head so as to decrease its thermal noise. The physical image of OMT is shown in Fig.\ref{Fig4-2}.

\subsection{Cryogenic unit}
\label{sect:Obs}

The cryogenic cooling unit basing on GM cycle \citep{Thirumaleshwar1986} is built to provide vacuum and cryogenic environment for low noise microwave parts in the frontend of receiver. It consists of a cryogenic dewar, a GM cold head and a helium compressor.

\begin{figure}
   \centering
   \renewcommand\thefigure{5}
   \includegraphics[width=14.0cm, angle=0]{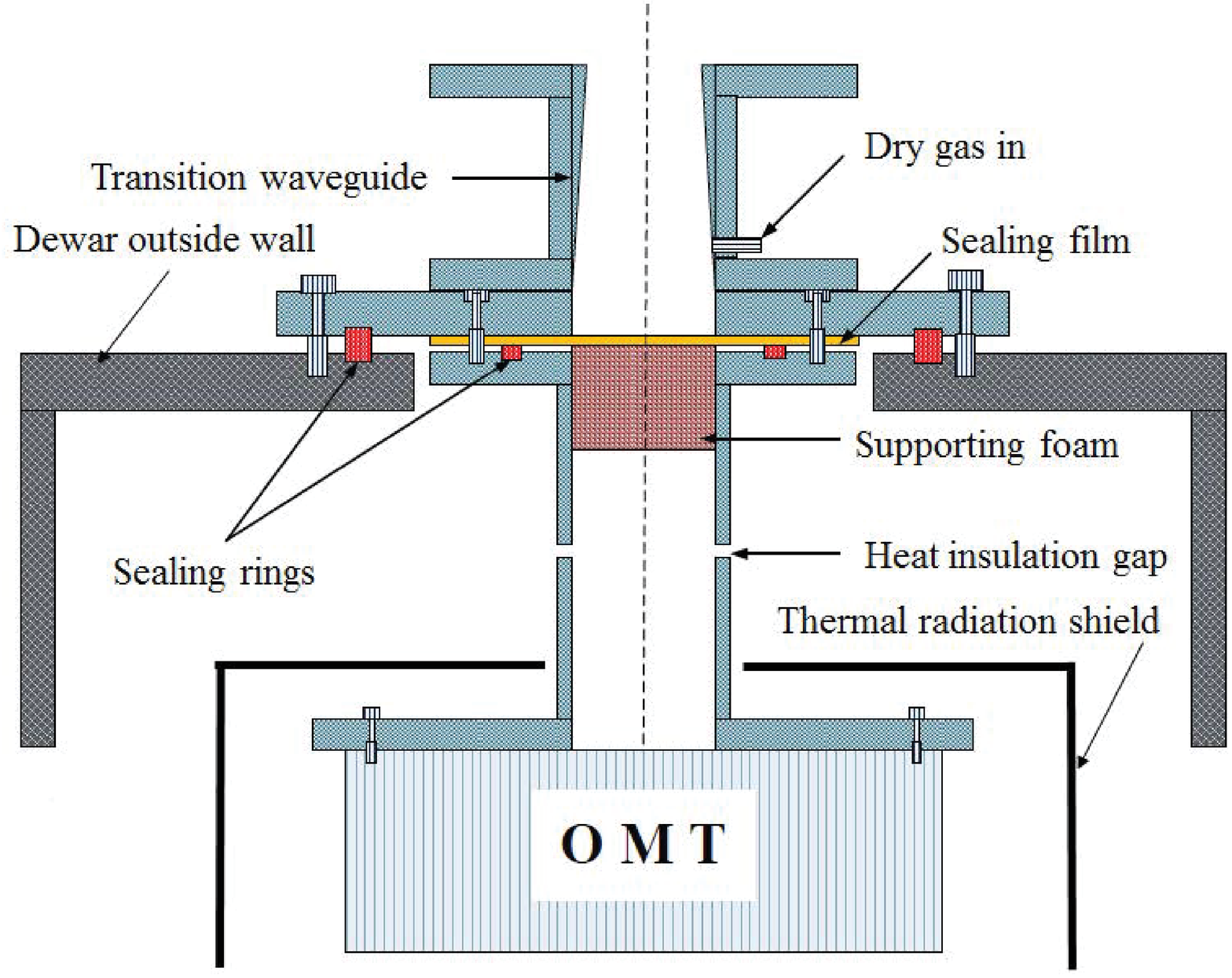}
   \caption{Sketch of cryogenic dewar of L-band receiver.   }
   \label{Fig5}
   \end{figure}

A sketch of cryogenic dewar of the proposed receiver is shown in Fig.\ref{Fig5}. Polyimide and Teflon bilayers films with Fluorine rubber sealing rings are used to seal vacuum window in view of rather large window diameter corresponding to OMT in L-band, meanwhile, a hard foam underneath is used to support the films at atmosphere pressure. The estimated insertion loss of the combination of bilayers films and foam will be less than 0.01 dB amounting to 0.7 K thermal noise to receiver system in L band. The heat insulation gap, vacuum degree of $10^{-5}$ Pa and thermal radiation shield will effectively eliminate the effects of conduction, convection and radiation of heats from dewar walls to cryogenic parts. Rectangular dewar structure is adopted, and the front panel can be removed as components inside dewar need to be repaired or replaced. Therefore, the sealing treatment between front panel and dewar is particularly important. A high quality Fluorine rubber sealing ring is applied, and the machining accuracy in panel, side walls of dewar and elongated slot are strictly controlled. The internal view of dewar can be seen in Fig.\ref{Fig2}.

\begin{figure}
   \centering
   \renewcommand\thefigure{6-1}
   \includegraphics[width=14.0cm, angle=0]{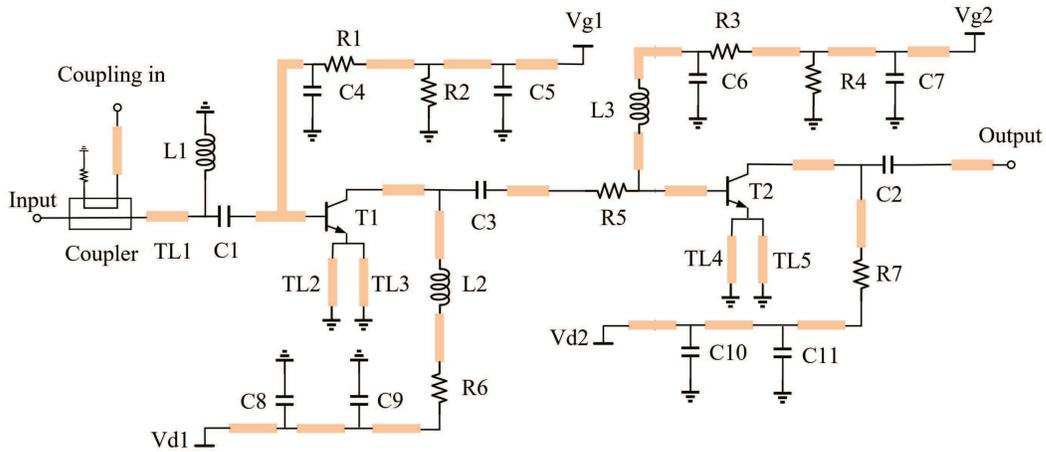}
   \caption{Schematic of coupling-LNA. Main functional elements of coupler and two HEMTs (T1 and T2) can be seen. Grid DC bias network of T1 consists of C4, R1, R2 and C5; drain DC bias network of T1 consists of L2, R6, C9 and C8; grid DC bias network of T2 consists of L3, C6, R3, R4 and C7; drain DC bias network of T2 consists of R7, C11 and C10; source negative feedback circuits of T1 and T2 consist of TL2,TL3,TL4 and TL5; input matching network mainly consists of TL1, L1 and C1; matching network between stages mainly consists of C3 and R5; output matching network mainly consists of C2 and R7.  }
   \label{Fig6-1}
   \end{figure}

\begin{figure}
   \centering
   \renewcommand\thefigure{6-2}
   \includegraphics[width=10.0cm, angle=0]{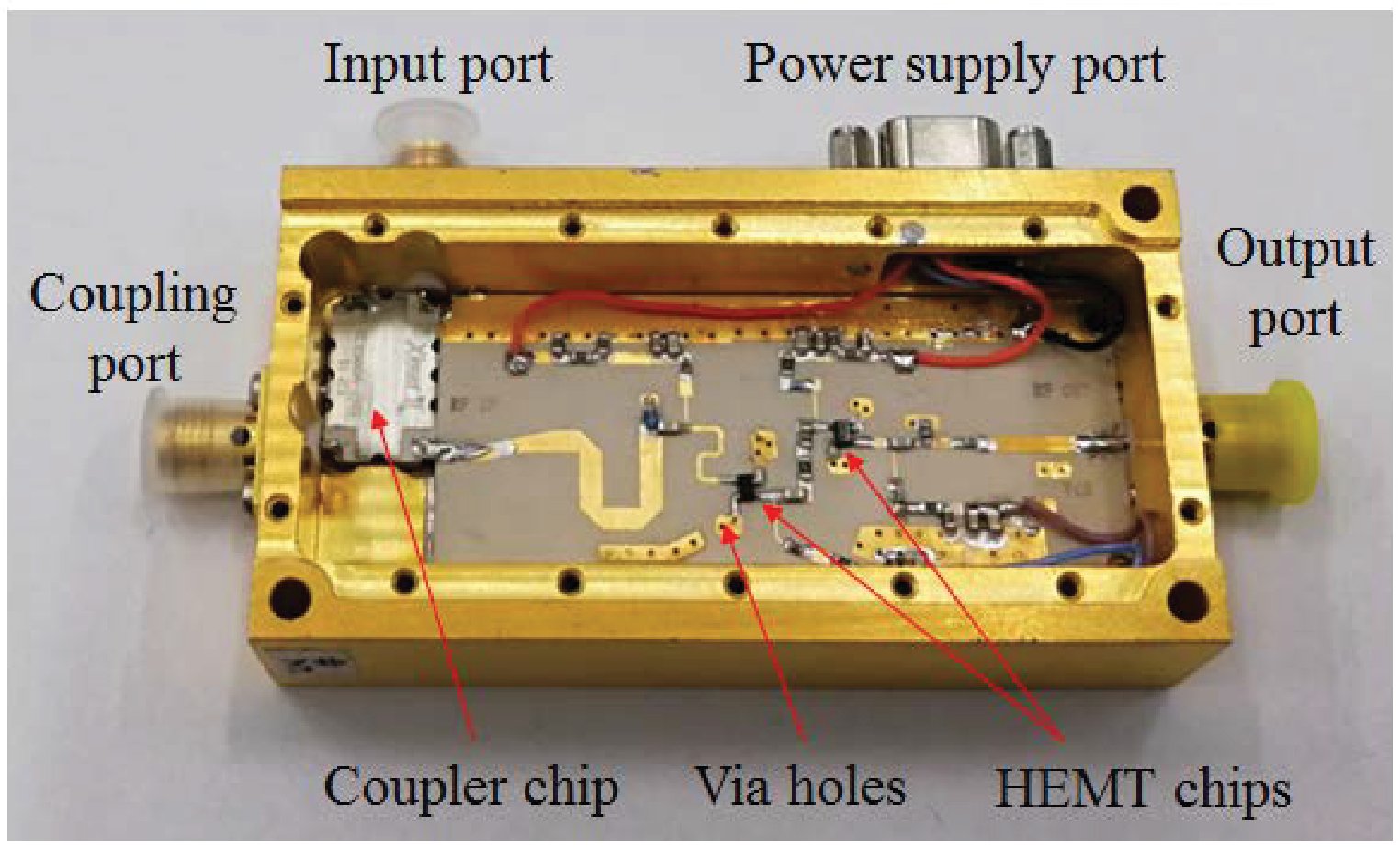}
   \caption{Completed L-band cryogenic Coupling-LNA with cover removed. Coupler chip, two HEMT chips, other lumped chips, double-layer microwave circuit board, microwave and power ports, and gold-plated brass box are shown.  The size of the shielded brass box is 70 mm x 37 mm x 17.6 mm. }
   \label{Fig6-2}
   \end{figure}

Behind OMT, two Coupling-LNAs are cooled to the physical temperature of 15 K by the second stage of cold head. The schematic of Coupling-LNA and internal view of fabricated Coupling-LNA are shown in Fig.\ref{Fig6-1} and Fig.\ref{Fig6-2}, respectively.

A 20 dB directional coupler chip of XC0450E-20S \cite{XC0450E-20S} from Anaren is integrated in the frontend of Coupling-LNA circuit for injecting calibration signals to main signal channels. On 15 K cold plate, 6 dB attenuator is employed between the coupling port of Coupling-LNA and cryogenic power splitter (4 dB power attenuation) to adjust calibrating noise power. So the total coupling coefficient is about 30 dB, which can suppress coupling noise to about 0.45 K as noise source is off. And two High Electron Mobility Transistor (HEMT) chips of ATF34143 \cite{ATF-34143} from Agilent are employed as amplification stages of circuits. The other selected lumped reactance elements - capacitors and inductors - are all the chips with high-quality factor for the purpose of reducing thermal noise, as well as low-loss microwave circuit board of Rogers RT6002 (0.0012 in loss tangent). Thermal stability for chips, circuit board and soldered dots are carefully concerned to ensure the reliability of circuit under cryogenic temperature.

Theoretically, all noises in LNA circuit can be broken down into two parts: passive thermal noise and transistor active noise. Passive noise derives from ohmic dissipation of all parts, which can be expressed as:
\begin{equation}
  T_{ohmic}= T_{p}(L-1)
\label{eq:3}
\end{equation}
Wherein, $T_p$ is physical temperature of elements, and $L$ is loss under $T_p$. Evidently, the passive noises are linearly dependent with physical temperature of elements. For transistors, besides the ohmic noise in pins, junctions and parasitic resistance, more noises derive from the shot noise and flicker noise of transistor's PN junctions. For a Field Effect Transistor (FET), the noise temperature $T_n$ can be expressed as \cite{Gallego1990}:
\begin{equation}
  T_{n} = T_{min}+T_{o}\frac{g_{n}}{R_{g}}|Z_{g}-Z_{gopt}|^{2}
\label{eq:4}
\end{equation}
Wherein, $T_o$ is standard temperature of 290 K, and $R_g$ is the practical generator resistance facing to FET. The four noise parameters of an FET at a certain DC bias - the available minimum noise temperature ($T_{min}$), noise conductance ($g_n$) and optimal source impedance ($R_{opt}$ + j$X_{opt}$ = $Z_{opt}$) - depend on operating frequency and ambient temperature. It can be seen from Equation (\ref{eq:4}) that the key design work for achieving low noise of FET is to form an output source impedance of $Z_g$ as close as possible to $Z_{opt}$ by an appropriate input matching circuit between 50 Ohm source and transistor. As a specific case, if $Z_g$ = $Z_{gopt}$, FET will achieve $T_{min}$. Generally, the noise parameters and resultant the noise temperature of FET are directly influenced by physical temperature of FET \cite{Pospieszalski1989}. Overall, whether passive thermal noise or FET active noise, good head-conduction connection from cooling head to chip elements is very important for achieving as low as possible noise.

Optimization of LNA requires the trade-off design among low noise, large return loss, high gain, unconditional stability and high dynamic range. As the utmost important elements affecting above performance of the proposed Coupling-LNA, microstrip lines of TL2, TL3, TL4 and TL5 (see Fig.\ref{Fig6-1}) constitute inductive negative feedback circuits in source electrodes of two HEMTs, and their basic roles are to improve stability of amplification circuit. In-depth optimization of TL2 and TL3 increases the real part of the source impedance achieving maximum delivery power, which is going to be closer to the source resistance achieving minimum noise temperature ($R_{opt}$). They are critical for realizing low noise matching while taking high delivery power matching into account.   Optimization of TL4 and TL5 helps to achieve good power matching between two HEMTs.  Two key points contribute to high dynamic performance of the Coupling-LNA: the selection of high dynamic characteristic transistor of ATF34143 and the design of positive (Vd1 and Vd2) and negative (Vg1 and Vg2) DC bias circuits. Fine tuning for HEMTs' grid voltages of Vg1 and Vg2 and drain voltages of Vd1 and Vd2 provides appropriate operating points required for high dynamic range of HEMTs. At the same time, they are also involved in the optimization of noise and power matching.

Heat produced by HEMT chips is mainly conducted to the cold head through upper metal layer of circuit board, via holes, lower metal layer and brass box in turn. Good thermal connections from brass box to chips ensure ultra-low physical temperature on chips. In 2017, we reported a cryogenic LNA with a similar HEMT chip of ATF54143 and achieved the noise temperature of 6 K \cite{Liu2017}. In this work, better heat conductivity treatment and more noise optimization focusing on this band are implemented to improve the noise temperature to 4 K even with an integrated coupler.

\subsection{Warm microwave unit}
\label{sect:Obs}

After the output of dewar, warm microwave unit is built for further signals processing. From Fig.\ref{Fig1} and Table \ref{T2}, amplifiers 1 with gain of 30 dB are used to further amplify signals; isolators with reverse isolation of 40 dB and forward loss of  0.6 dB are used for improving power matching of chains; bandpass filters with loss of 1 dB are for shaping the excepted frequency band of 1.2 GHz - 1.8 GHz; remote controlled digital attenuators with intrinsic loss of 1.5 dB can provide a gain adjustment range of 30 dB with the steps of 1 dB, 5 dB and 10 dB to cope with different interference signal powers; RF detectors pick up signals' full band power via splitters to check channels situation of receiver; the other output of splitters are connected to amplifiers 2 with gain of 20 dB and output 1 dB compression power of +15 dBm, and high dynamic range is necessary for the post amplifiers to keep linearity of receiver. RF over fiber optical transmitters are used for delivering the signals from receiver cabin to laboratory on the ground via about 3 km high-stability bendable optical fibers \cite{Liu2017optical}. In addition, calibrating signals unit provide noise power calibrating and phase calibrating function, which mainly includes a solid-state noise source and a comb signals calibrating source in a 40 degree Celsius thermostat, power splitters and combiners, and microwave switches. Two levels of noise temperatures of 1 K and 10 K are provided for different calibration applications of telescope.

\section{MODULES AND SYSTEM MEASUREMENT}
\label{sect:Obs}

\subsection{Feed network}
\label{sect:Obs}

\begin{figure}
   \centering
   \renewcommand\thefigure{7}
   \includegraphics[width=10.0cm, angle=0]{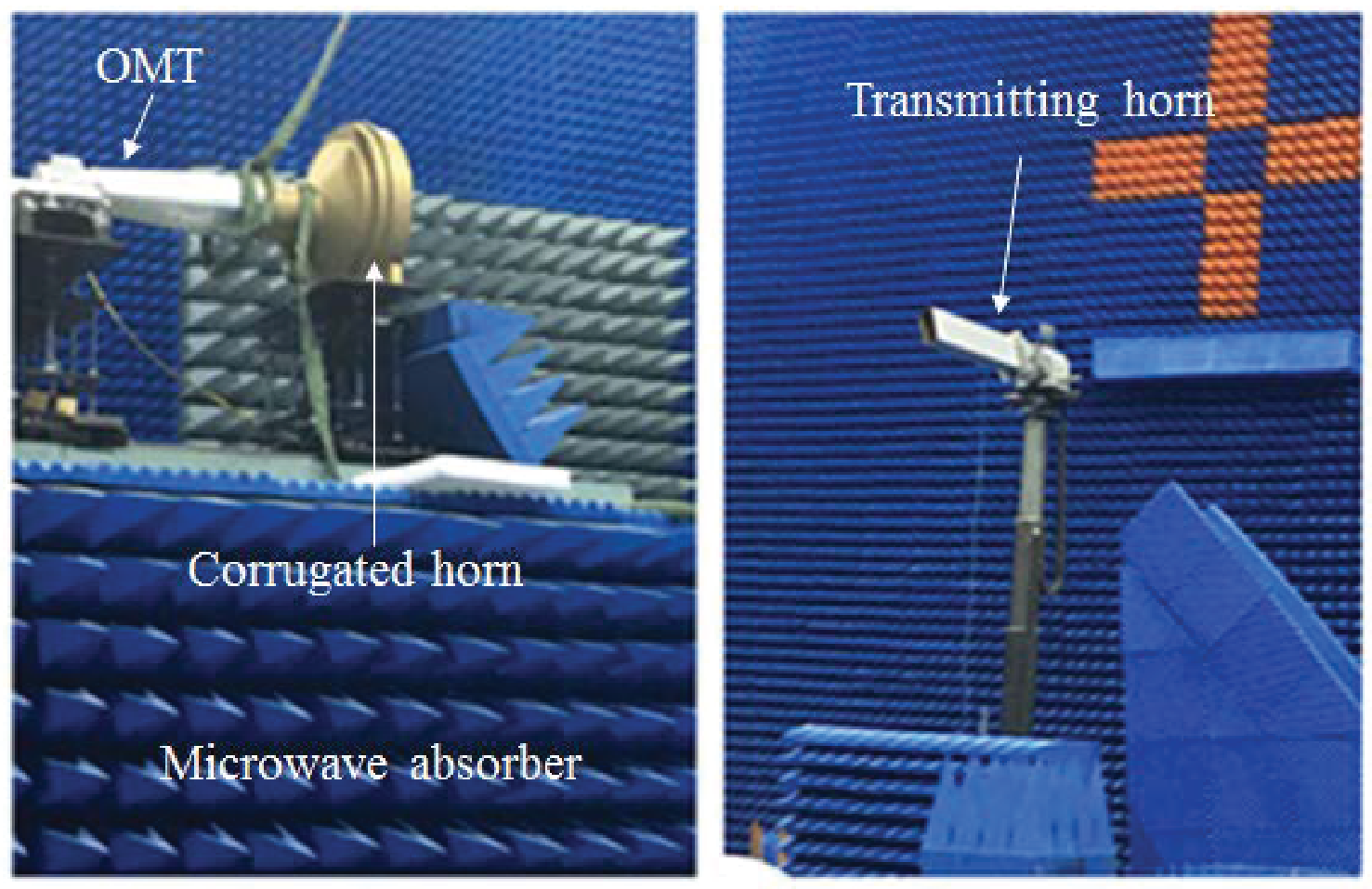}
   \caption{Radiation patterns measurement setup of feed network in 10 m microwave anechoic room. Corrugated horn and OMT under test, transmitting horn and microwave absorbers can be seen.}
   \label{Fig7}
   \end{figure}

\begin{figure}
   \centering
   \renewcommand\thefigure{8}
   \includegraphics[width=14.0cm, angle=0]{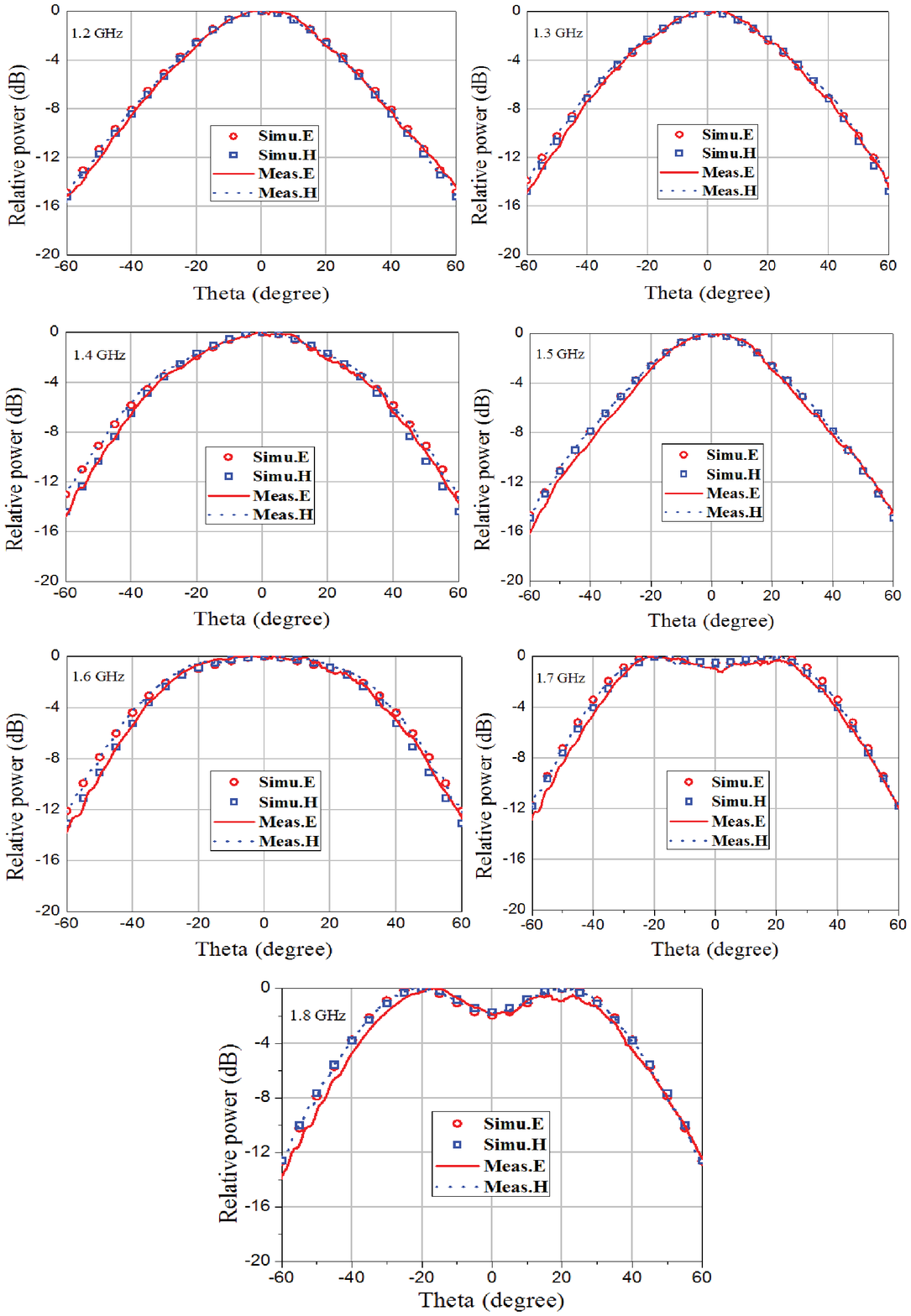}
   \caption{Measured and simulated E-plane and H-plane normalized radiation patterns of feed network at 1.2 GHz, 1.3 GHz, 1.4 GHz, 1.5 GHz,1.6 GHz, 1.7 GHz and 1.8 GHz. }
   \label{Fig8}
   \end{figure}

The radiation patterns of feed network were measured in a microwave anechoic room (shown in Fig.\ref{Fig7}) by utilizing a transmitting horn and Agilent E5071C Vector Network Analyzer (VNA). The measured and simulated radiation patterns of E-plane and H-plane are shown in Fig.\ref{Fig8}. It can be seen that the measured results fit very well with the simulated results no matter E-plane or H-plane in whole passband. Which demonstrates good simulation precision and machining accuracy. By comparing E-plane with H-plane of pattern at each frequency point, it can be found that the equalization of patterns is also excellent, especially in low frequency band.

\begin{figure}
   \centering
   \renewcommand\thefigure{9}
   \includegraphics[width=14.0cm, angle=0]{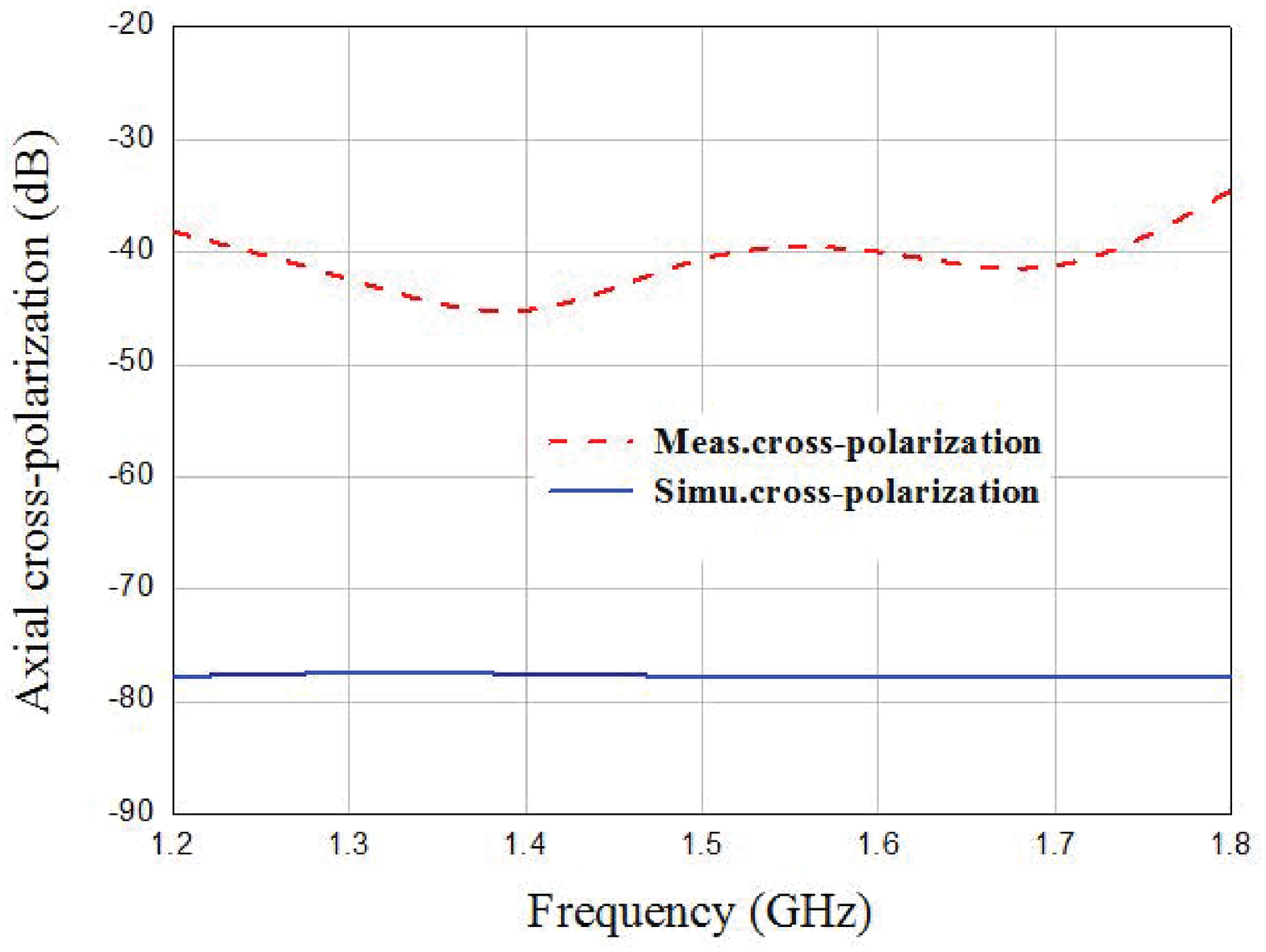}
   \caption{Simulated and measured axial cross-polarization level of the corrugated horn.}
   \label{Fig9}
   \end{figure}

\begin{figure}
   \centering
   \renewcommand\thefigure{10}
   \includegraphics[width=14.0cm, angle=0]{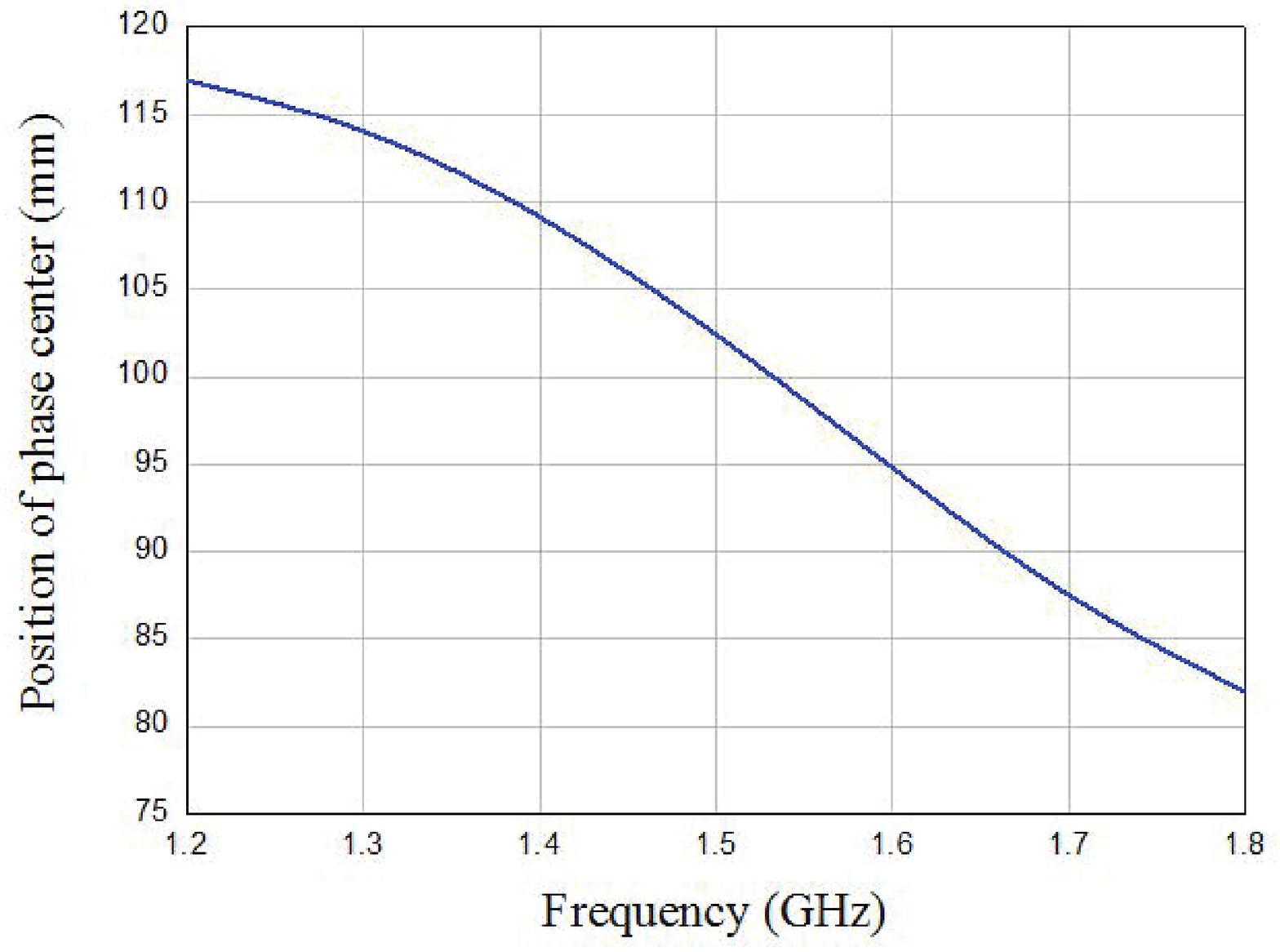}
   \caption{Simulated distances from phase center to aperture plane of the corrugated horn.}
   \label{Fig10}
   \end{figure}

The corrugated slots effectively eliminated the E-plane edge diffraction and then improved the equalization of E-plane and H-plane patterns, and the deviations are less than 1 degree in passband. Meanwhile, it is met well that 10 dB edge illumination are in about 110 degrees flare angle in whole band. In addition, the simulated and measured axial cross-polarization level of horn are shown in Fig.\ref{Fig9}. For a rotational symmetric and almost ideal simulation model of horn, the simulated cross-polarization levels are always quite low, and machining error for actual horn deteriorated cross-polarization level, while it is still lower than -34 dB over passband. The simulated phase center positions varying with frequencies are shown in Fig.\ref{Fig10}, and the maximum deviation over passband is 0.175$\lambda_c$ and the optimization of dish aperture efficiency varying with phase centers over passband shows the distance to horn aperture plane of 85.84 mm is the optimal phase center corresponding to the maximum deviation over passband of 0.156$\lambda_c$ for practical application.

\begin{figure}
   \centering
   \renewcommand\thefigure{11}
   \includegraphics[width=14.0cm, angle=0]{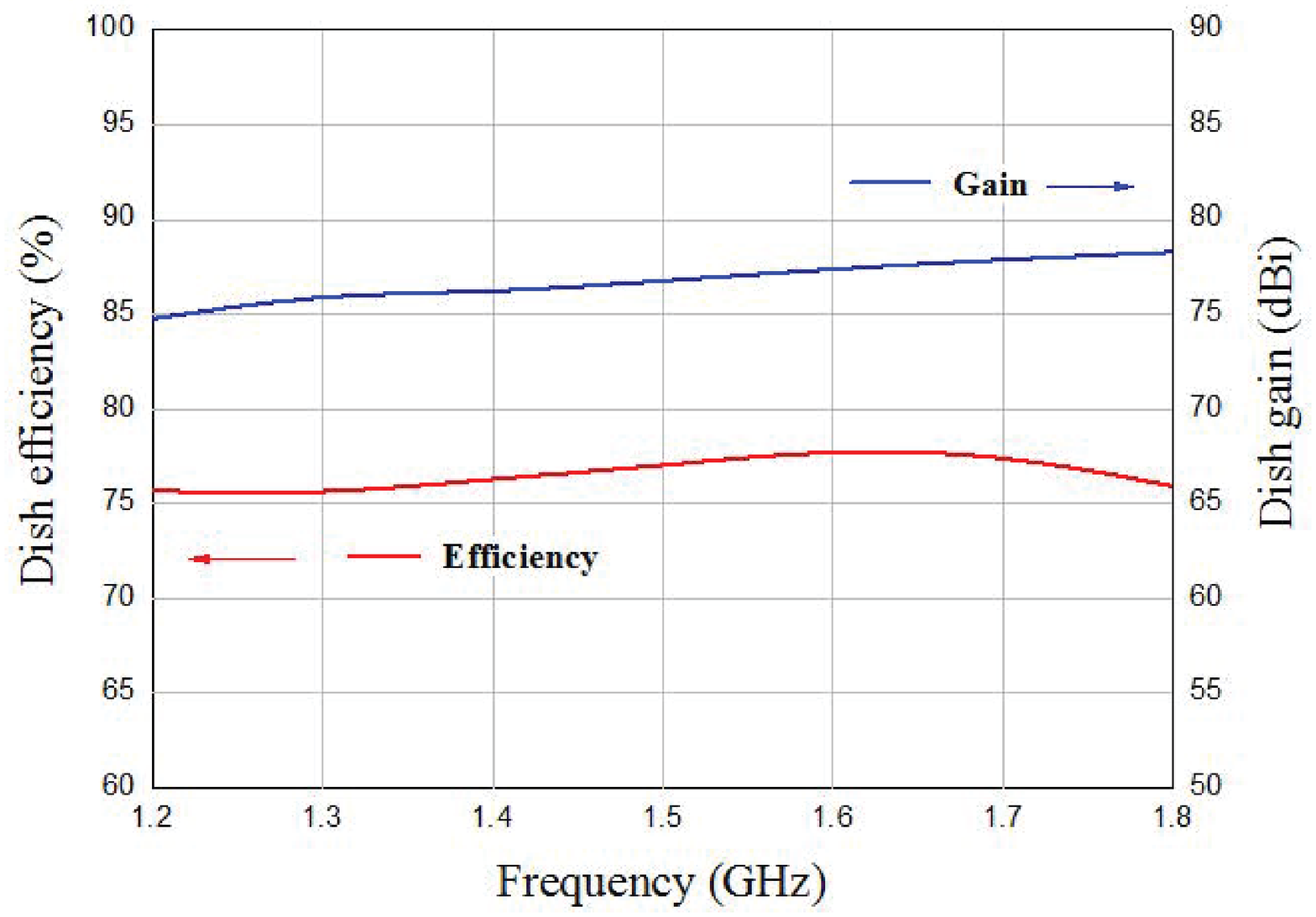}
   \caption{Simulated dish efficiency and gain with measured radiation patterns of horn and ideal FAST reflector model based on physical optics method.}
   \label{Fig11}
   \end{figure}

By using the measured horn's radiation pattern and continuous FAST reflector model, the simulation of dish aperture efficiency and gain based on physical optics method were implemented in an antenna simulator \cite{Grasp}. Dish aperture efficiency more than 75\% and dish gain larger than 74 dBi were achieved (see Fig.\ref{Fig11}). The actual dish efficiency and gain will be lower than that in above simulation due to the existence of gaps and holes on reflectors, and deviation from ideal paraboloid of the reflectors. It is difficult to implement a more precise full-wave electromagnetic simulation with actual reflector model for an electrically object as large as FAST with 500 m dimensions at L band. Experimental test for dish efficiency is necessary with the proposed receiver installed on FAST when the existing observation missions with 19-beam receiver are finished in the future.

\begin{figure}
   \centering
   \renewcommand\thefigure{12}
   \includegraphics[width=12.0cm, angle=0]{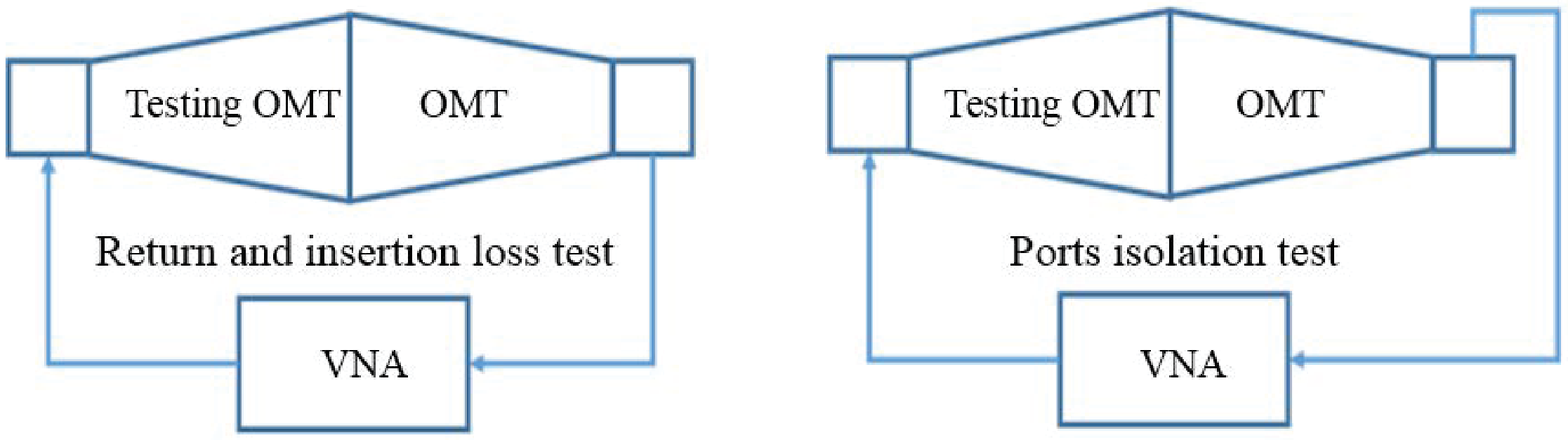}
   \caption{Two ports return loss, insertion loss (left) and ports polarization isolation (right) of OMT test configurations. The identical polarization ports are connected by VNA in left configuration; the vertical polarization ports are connected by VNA in right.}
   \label{Fig12}
   \end{figure}

\begin{figure}
   \centering
   \renewcommand\thefigure{13}
   \includegraphics[width=14.0cm, angle=0]{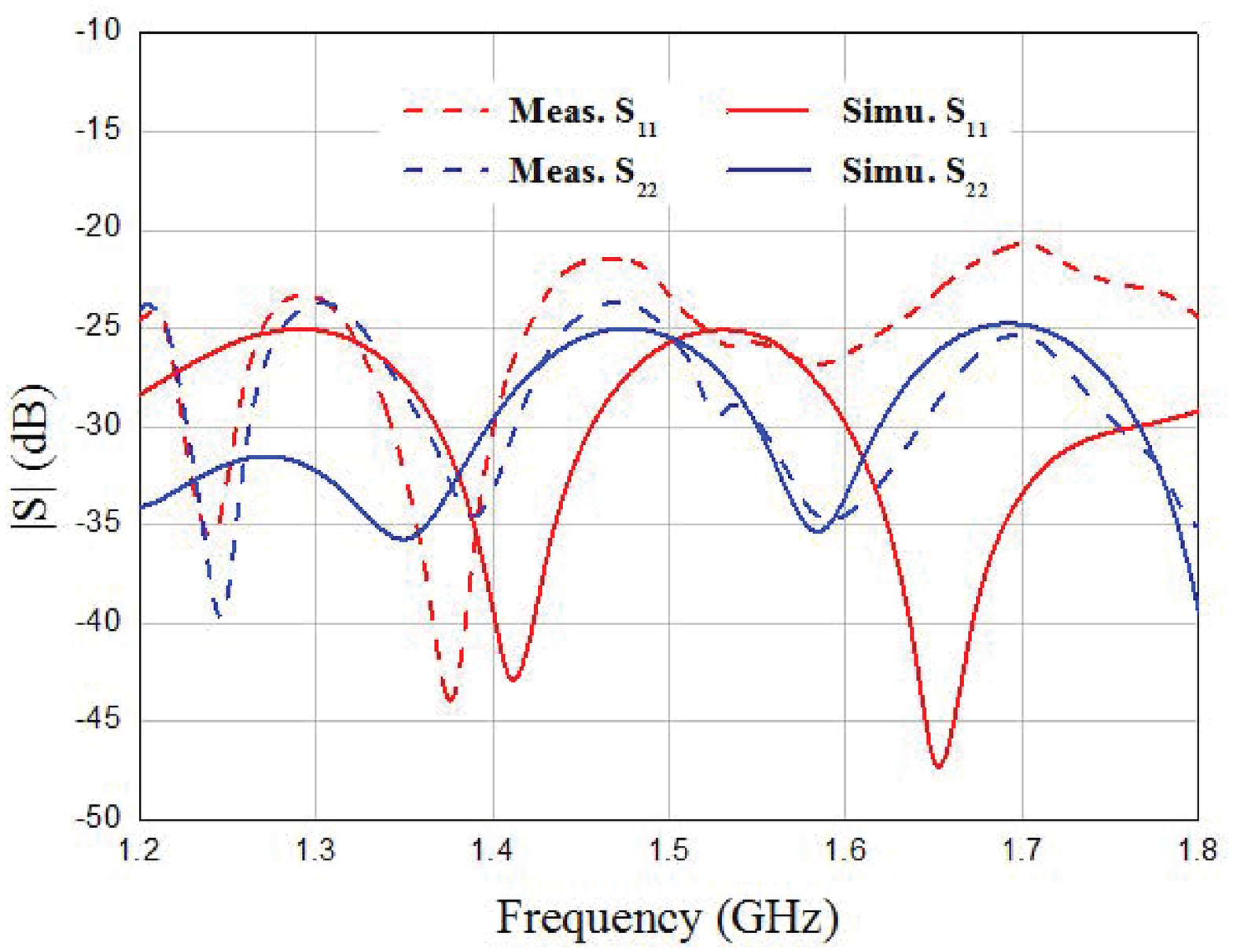}
   \caption{Measured and simulated ${\rm S_{11}}$ and ${\rm S_{22}}$ of OMT.}
   \label{Fig13}
   \end{figure}

\begin{figure}
   \centering
   \renewcommand\thefigure{14}
   \includegraphics[width=14.0cm, angle=0]{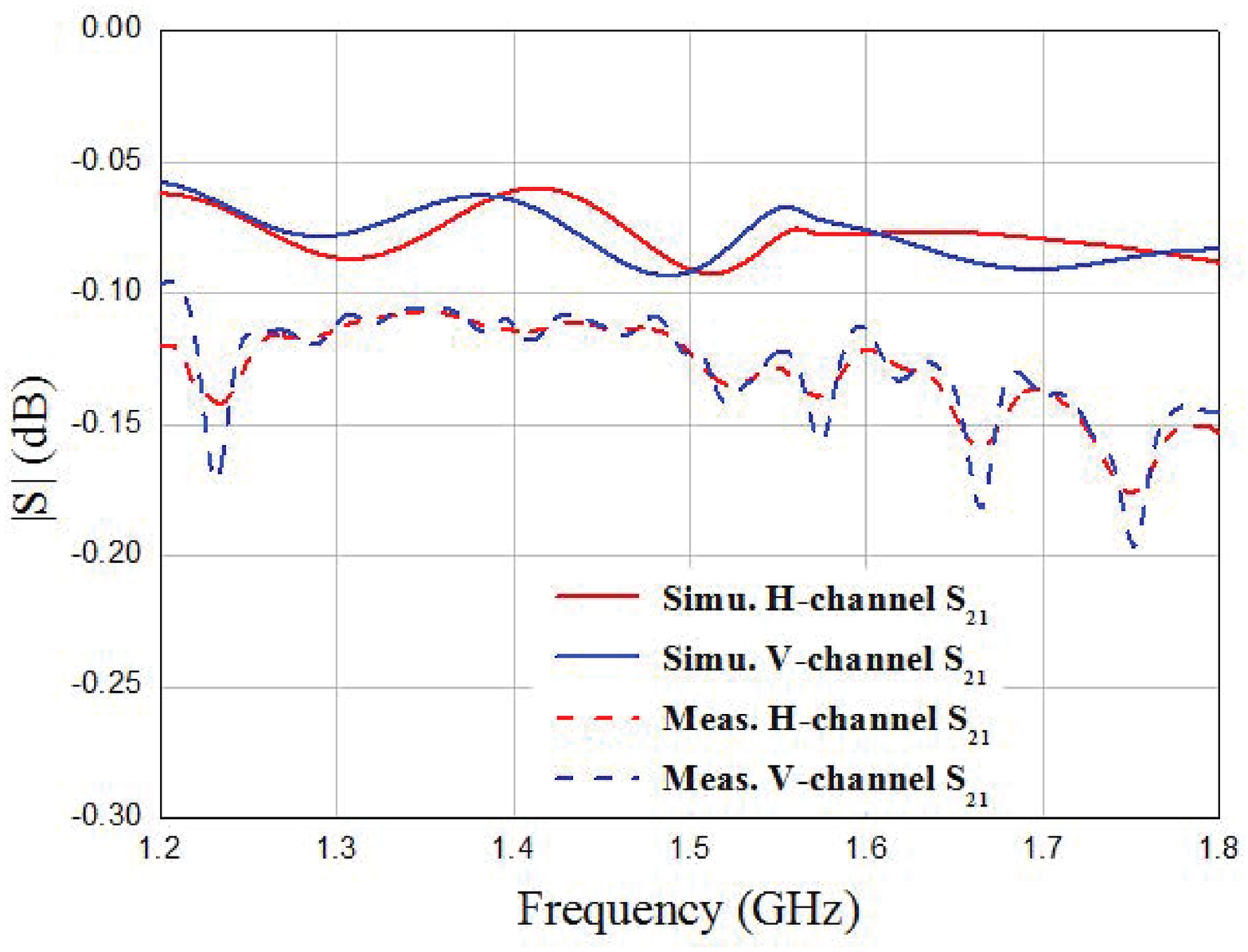}
   \caption{Measured values of ${\rm S_{21}}$ by connecting back to back two identical OMT units. Measured amount in dB is divided by 2 to get ${\rm S_{21}}$ values of an individual OMT.}
   \label{Fig14}
   \end{figure}

\begin{figure}
   \centering
   \renewcommand\thefigure{15}
   \includegraphics[width=14.0cm, angle=0]{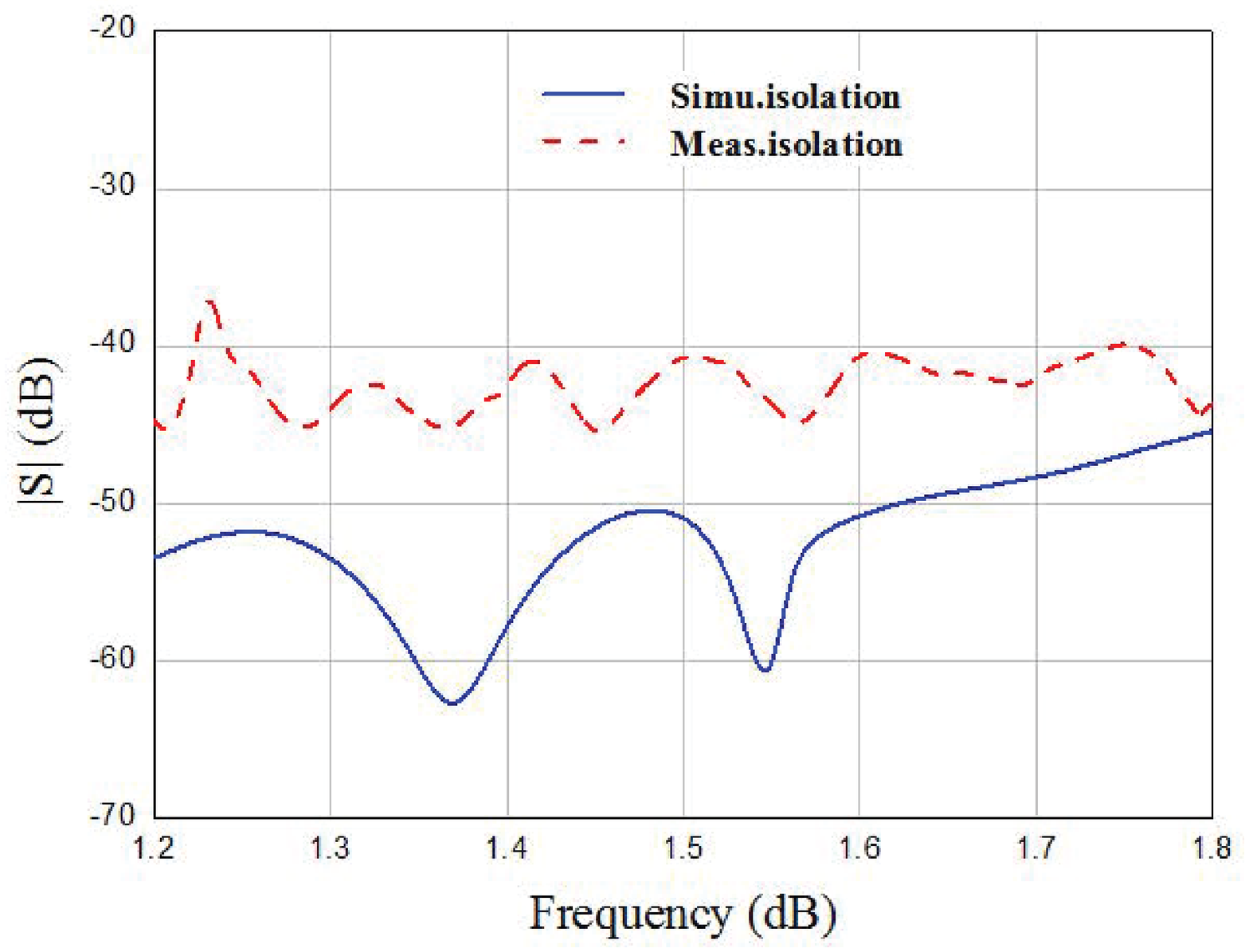}
   \caption{Measured and simulated results of ports polarization isolation of OMT.}
   \label{Fig15}
   \end{figure}

The S-parameters of OMT were measured with VNA by connecting back to back two identical OMT units and keeping the ridges with same polarizations aligned. The test configuration is shown in Fig.\ref{Fig12}. Two ports return loss, insertion loss, and ports polarization isolation can be figured out in this measurement. The measured results of ${\rm S_{11}}$ and ${\rm S_{22}}$ as well as the corresponding simulated results are shown in Fig.\ref{Fig13}. The variation tendency of measured curves is generally consistent with the simulated curves, except for ${\rm S_{11}}$ at 1.6 GHz to 1.8 GHz. It can be seen that two ports return loss are all greater than 20 dB from 1.2 GHz to 1.8 GHz, which are good for power matching with the following components and then eliminating passband ripples caused by standing waves. The measured and simulated results of ${\rm S_{21}}$ for OMT are shown in Fig.\ref{Fig14}. It shows that the insertion loss is less than 0.2 dB from 1.2 GHz to 1.8 GHz, and the average insertion loss is about 0.12 dB, which will introduce less than 2 K noise to receiver system. It is noticed that there are 4 singular points in the measured ${\rm S_{21}}$ curves caused by signal resonance in back to back setup, which won't introduce the thermal noise as much as that corresponding to the nominal insertion loss values. Compared with simulation, the measured insertion loss is generally worse, while they are still under expected. It is difficult to establish a simulation model which can accurately enough representing the actual loss of OMT with non-perfect surface. The measured ports polarization isolation as well as the simulated curves are shown in Fig.\ref{Fig15}. Except for a hump with 37 dB resonance peak at 1.23 GHz, the isolation is greater than 40 dB, which can evidently eliminate cross talk between two channels of receiver.

\subsection{Cryogenic Coupling-LNA}
\label{sect:Obs}

\begin{figure}
   \centering
   \renewcommand\thefigure{16}
   \includegraphics[width=14.0cm, angle=0]{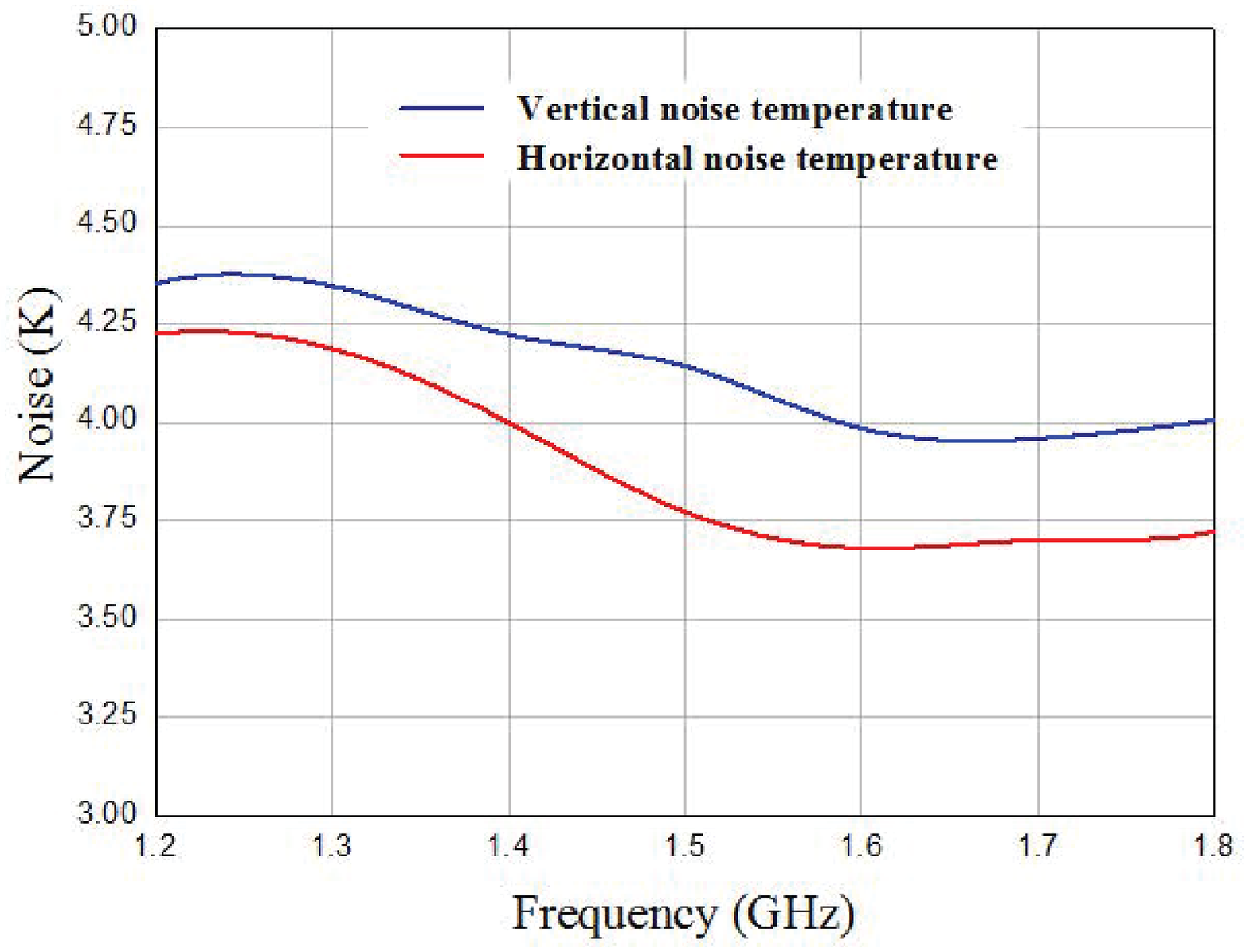}
   \caption{Measured noise temperature of Coupling-LNAs in vertical and horizontal polarization of receiver.}
   \label{Fig16}
   \end{figure}

The noise temperature of Coupling-LNA of the proposed receiver at the physical temperature of 15 K were measured by cryogenic noise testing system basing on Variable Load Temperature method \cite{Zhang2010}. The measured results are shown in Fig.\ref{Fig16}. The noise level and variation tendency with frequencies for two polarizations' Coupling-LNAs are very consistent. The mean in passband of noise temperature for horizontal and vertical polarization of Coupling-LNA are 3.9 K and 4.1 K, respectively. It shows excellent noise performance, especially considering a coupler is integrated in the front.

\begin{figure}
   \centering
   \renewcommand\thefigure{17}
   \includegraphics[width=14.0cm, angle=0]{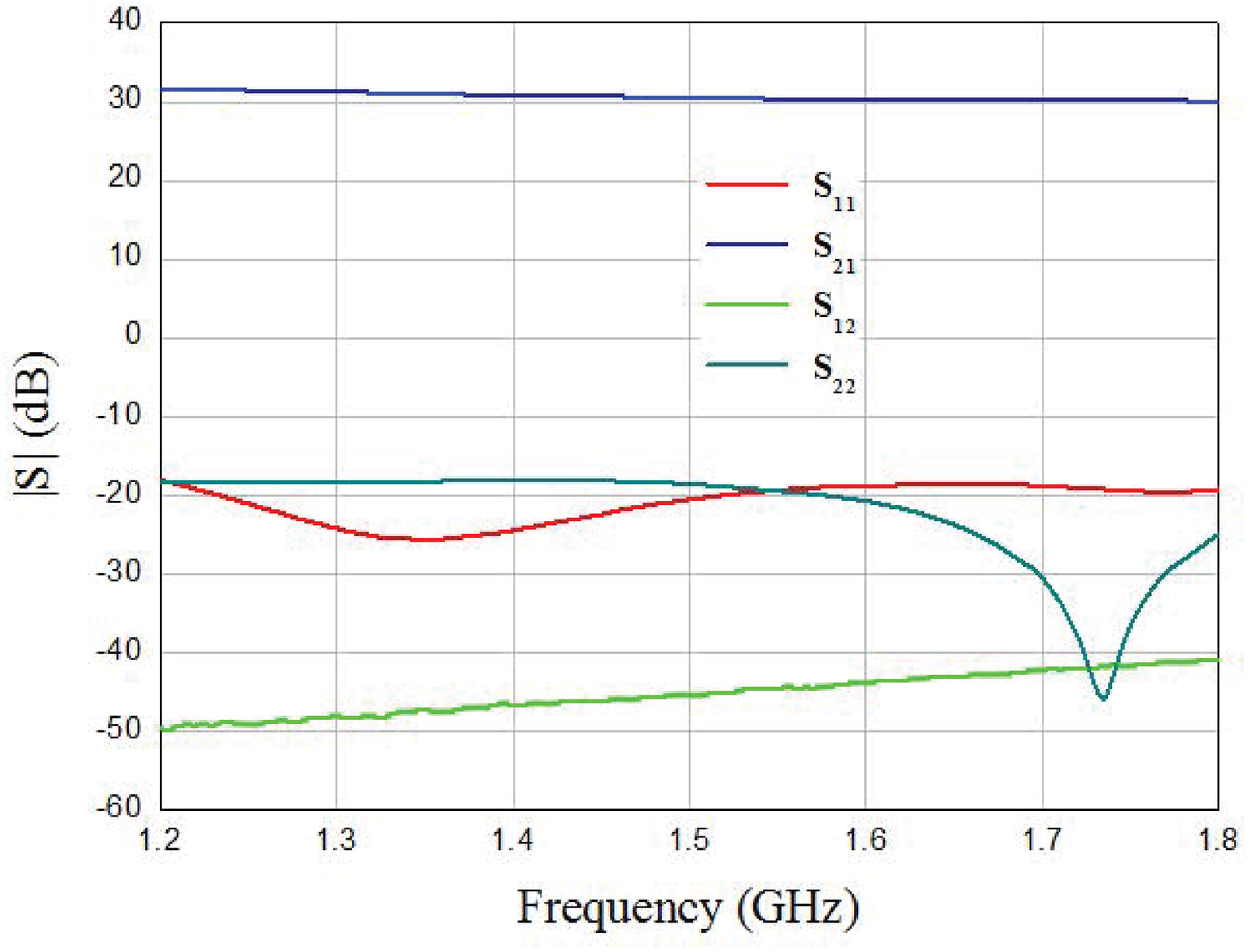}
   \caption{Measured main channel S-parameters of Coupling-LNA at 15 K.}
   \label{Fig17}
   \end{figure}

The measured main channel S-parameters of Coupling-LNA at 15 K are shown in Fig.\ref{Fig17}. ${\rm S_{11}}$ less than -18 dB in whole passband and less than -20 dB in most bands of 1.2 GHz to 1.8 GHz were achieved. Compared with input return loss of the other available cryogenic LNA without coupling function, such as 10 dB of CITLF3 and less than 10 dB of LNC0.6-2A, the proposed Coupling-LNA has obvious advantages in input matching. ${\rm S_{22}}$ is also less than -17 dB and very close to -20 dB from 1.2 GHz to 1.8 GHz. In view of the trade-off design between low noise and low input reflection has to be made, this is an LNA with good application values. The main channel (Input port to Output port, Fig.\ref{Fig6-2}) gain of about 31 dB was achieved, and gain flatness is better than +/-0.8 dB.

\begin{figure}
   \centering
   \renewcommand\thefigure{18}
   \includegraphics[width=14.0cm, angle=0]{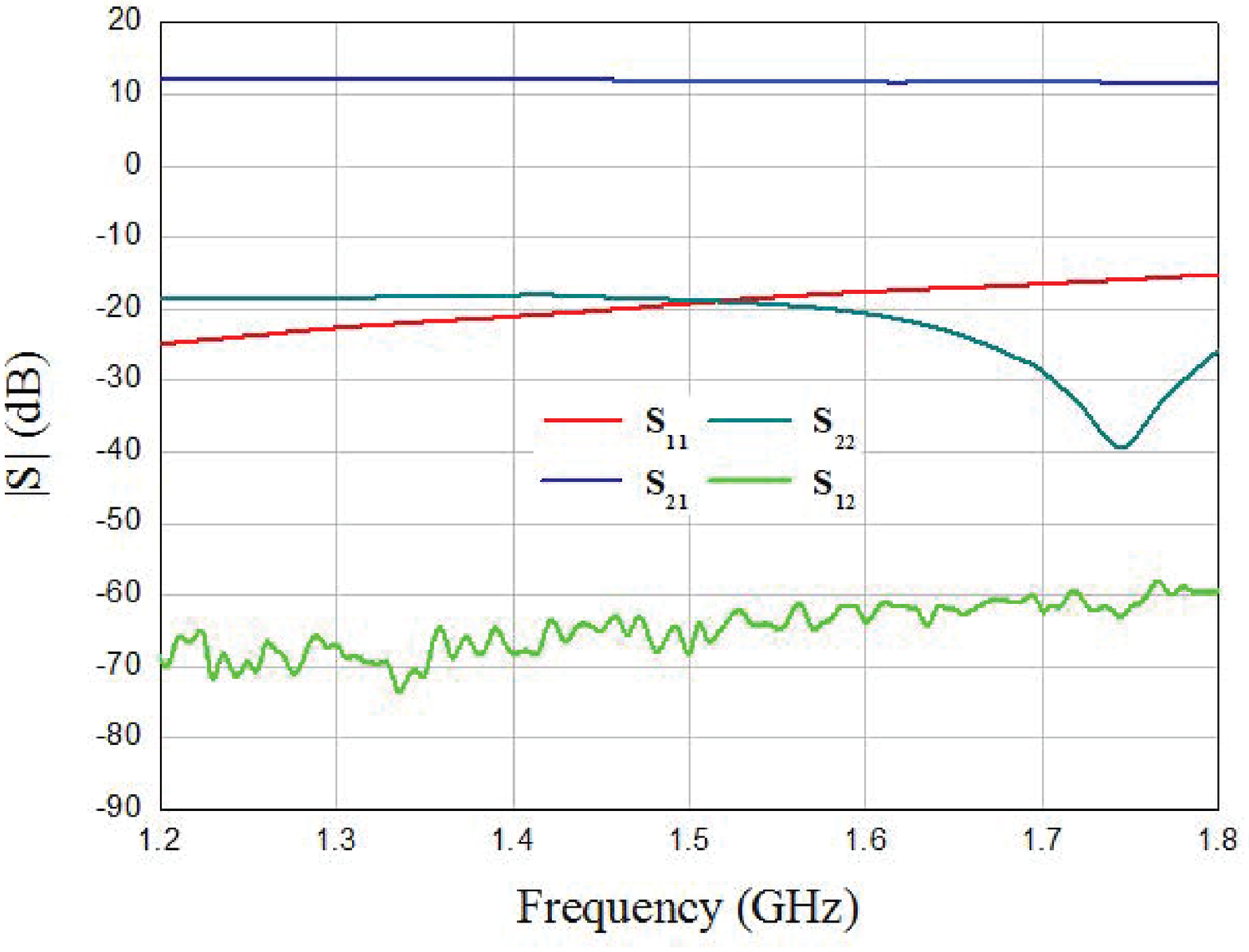}
   \caption{Measured coupling channel S-parameters of Coupling-LNA at 15 K.}
   \label{Fig18}
   \end{figure}

The S-parameters of noise coupling channel (Coupling port to Output port, Fig.\ref{Fig6-2}) of Coupling-LNA at 15 K was also measured, and results are shown in Fig.\ref{Fig18}. Gain is about 11 dB, which demonstrates the coupling coefficient of 20 dB. ${\rm S_{11}}$ is less than -15 dB and ${\rm S_{22}}$ is the same with that in main channel measurement.

The power-sweeping test shows the output 1 dB compressed power of Coupling-LNA is +5 dBm. Which is high enough for astronomical applications, and mainly profiting from high dynamic characteristic of HEMT ATF34143 and good bias design.

\subsection{Receiver system}
\label{sect:Obs}

\begin{figure}
   \centering
   \renewcommand\thefigure{19}
   \includegraphics[width=14.0cm, angle=0]{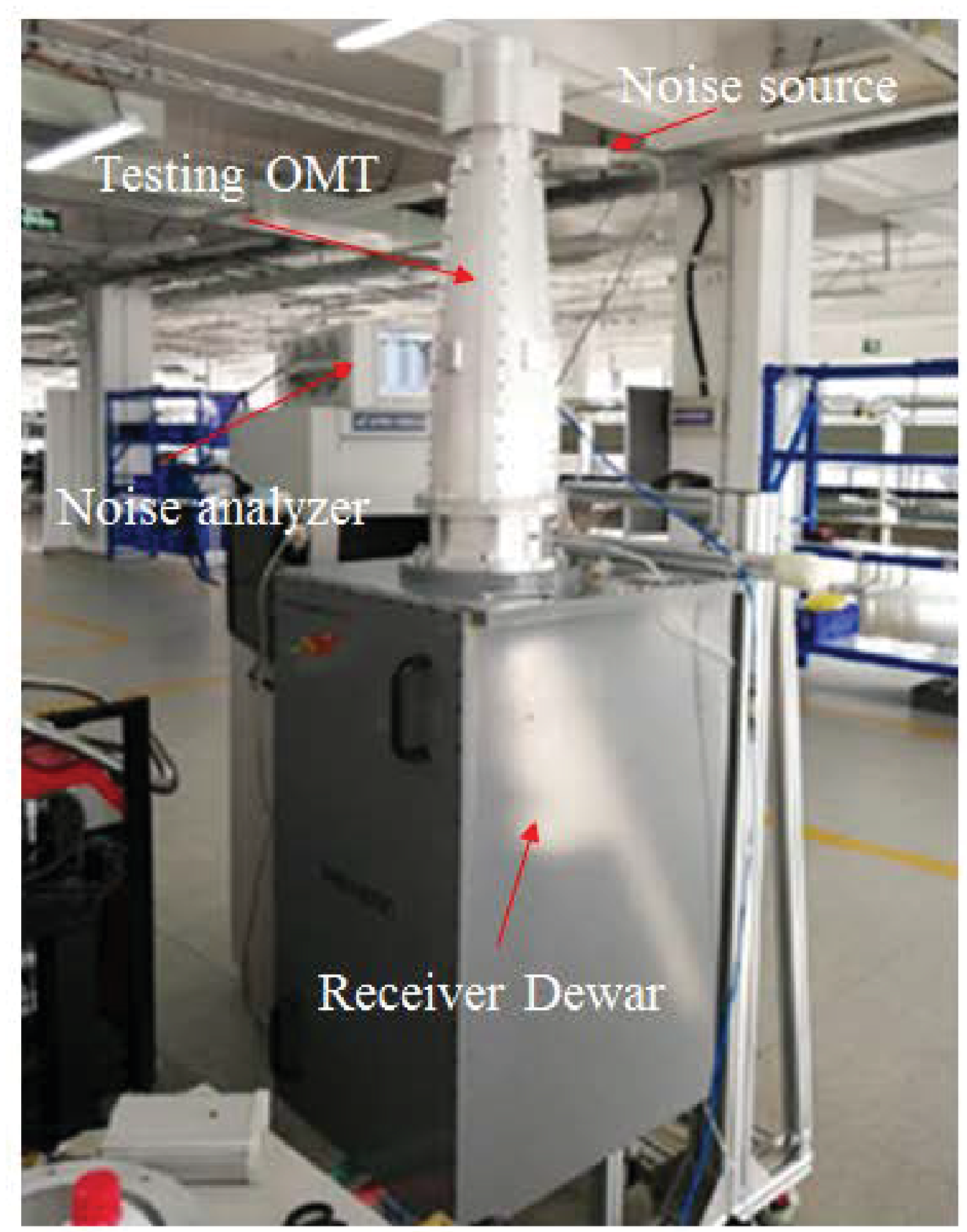}
   \caption{Measurement setup of receiver noise. Receiver with horn removed, testing OMT, noise figure analyzer and noise source can be seen.}
   \label{Fig19}
   \end{figure}

\begin{figure}
   \centering
   \renewcommand\thefigure{20-1}
   \includegraphics[width=14.0cm, angle=0]{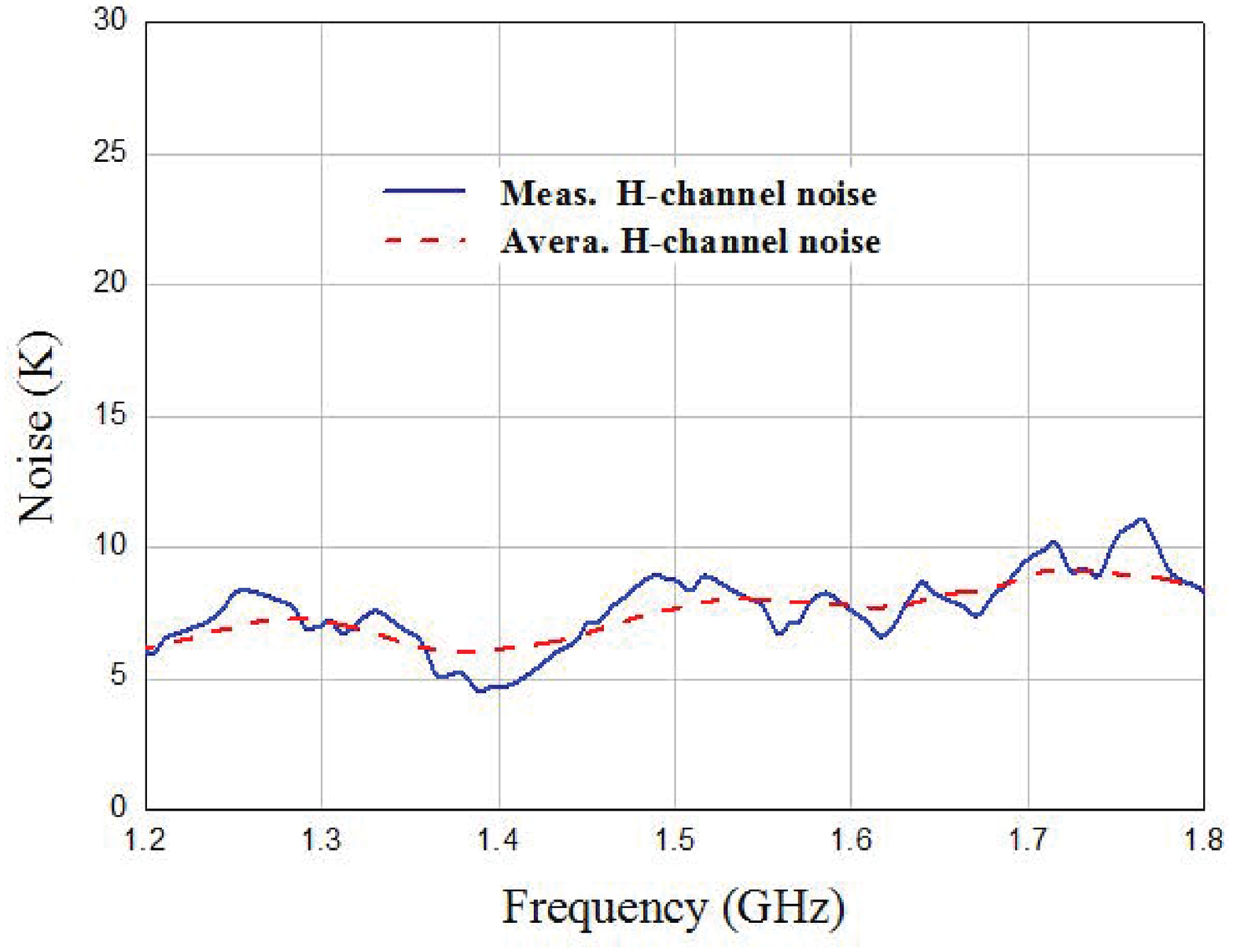}
   \caption{Noise temperature of horizontal polarization of the proposed receiver (blue line). Moving average with 10 adjacent frequency points for the measured data were implemented to achieve average noise level hiding in the time invariant interference (red dashed line).}
   \label{Fig20-1}
   \end{figure}

\begin{figure}
   \centering
   \renewcommand\thefigure{20-2}
   \includegraphics[width=14.0cm, angle=0]{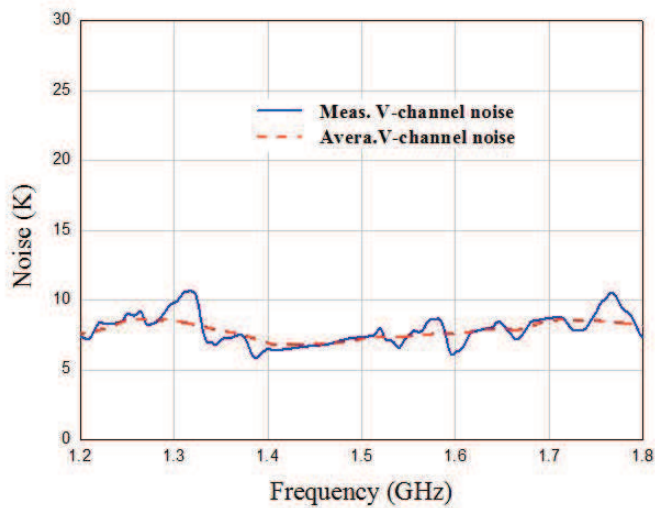}
   \caption{Noise temperature of vertical polarization of the proposed receiver (blue line).  Moving average with 10 adjacent frequency points for the measured data were implemented to achieve average noise level hiding in the time invariant interference (red dashed line).}
   \label{Fig20-2}
   \end{figure}

The noise temperature of receiver system was measured by Agilent 8974A noise figure analyzer with "smart" noise source of N4001A. The testing OMT are employed to transduce SMA output of N4001A to circular waveguide input of dewar, and the thermal noise introduced by the testing OMT are removed by utilizing the data of insertion loss presented in Section 3.1. The measurement setup is shown in Fig.\ref{Fig19}. The measured noise temperature of receiver system is shown in Fig.\ref{Fig20-1} and Fig.\ref{Fig20-2}. 10 times trace average for the measured noise is set in noise figure analyzer to eliminate time-varying interference. The average in band (1.2 GHz to 1.8 GHz) of noise temperature for horizontal and vertical polarization are 7.74 K and 7.85 K referred to vacuum window of dewar, respectively. This shows that it is a very sensitive receiver for dark astronomical source searching.

\begin{figure}
   \centering
   \renewcommand\thefigure{21}
   \includegraphics[width=14.0cm, angle=0]{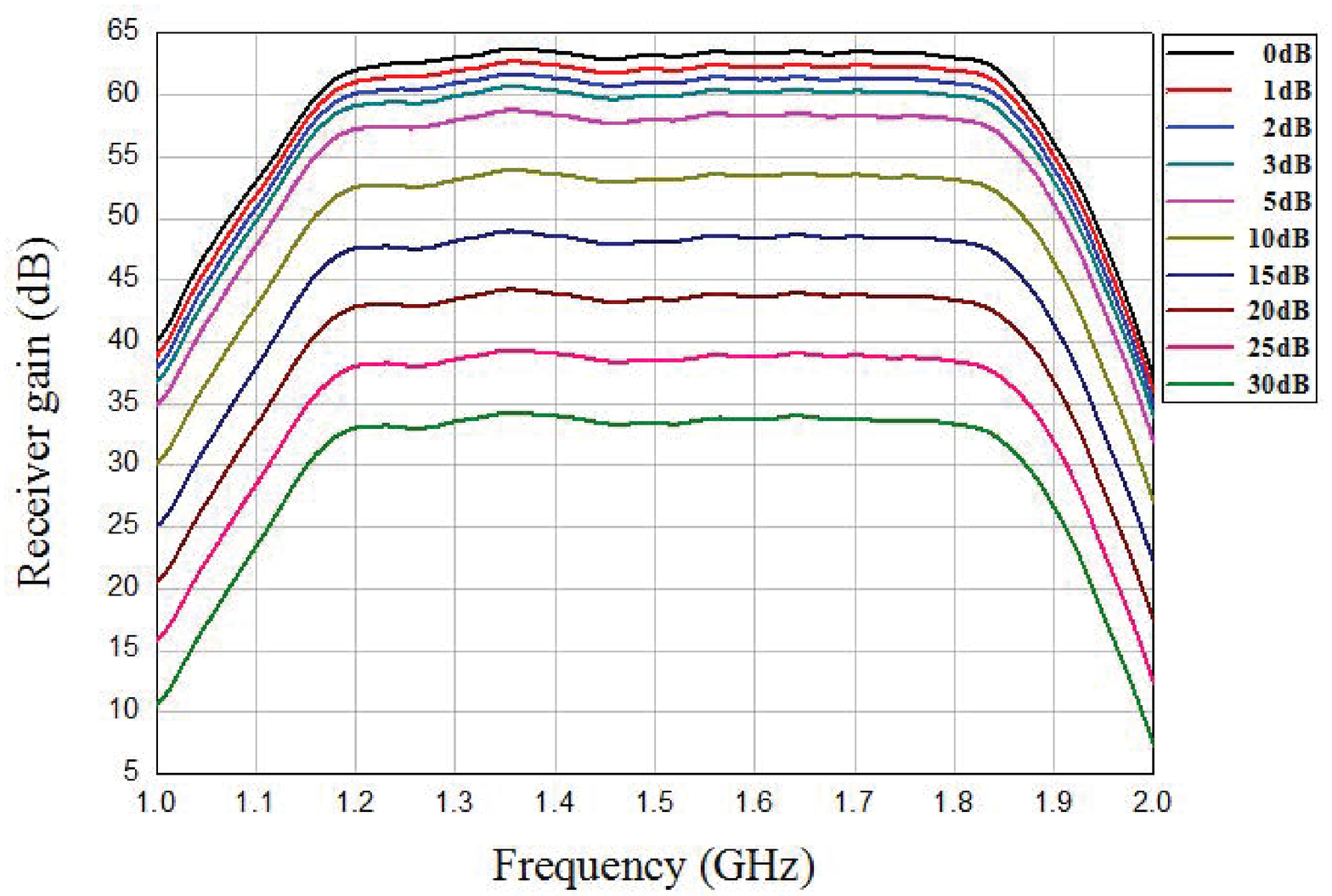}
   \caption{Frequency responses of horizontal polarization of the proposed receiver under 10 different gain levels.}
   \label{Fig21}
   \end{figure}

Keysight N9010B signal analyzer was used to test receiver's passband response with different gain levels by remote controlled digital attenuators. Digital attenuators can provide 30 dB gain adjustment range with the steps of 1 dB, 5 dB and 10 dB. The measured response with 10 gain levels of horizontal polarization are shown in Fig.\ref{Fig21} and the response in vertical polarization is almost the same. The ladder-shaped responses meet design requirement and the response ripples in passband is less than +/-1 dB.

\section{SUMMARY}
\label{sect:Obs}

We have developed an ultra-low noise cryogenic L-band receiver for FAST telescope and described its main performance. This receiver covers the frequency band from 1.2 GHz to 1.8 GHz. Novel cryogenic Coupling-LNA with the noise temperature of 4 K at the physical temperature of 15 K was developed for this receiver, and the noise temperature of the receiver system is below 9 K referred to feed aperture plane. A corrugated horn as well as a quad-ridge OMT were fabricated. And good scattering and radiation performance, and resultant dish efficiency of higher than 75\% were achieved. This is a very sensitive receiver compared with the other L-band receivers worldwide and will remarkably improve the detecting capacity of dark astronomical sources for FAST telescope in L band.

\begin{acknowledgements}

this work was supported by National Key R\&D Program of China (2018YFE0202900) and National Natural Science Foundation of China (U1831110).

\end{acknowledgements}


\end{document}